\documentclass[11pt,english]{article}
\usepackage{graphicx}
\usepackage{xcolor}
\usepackage{amstext}
\usepackage{caption}
\usepackage{etoolbox}
\usepackage{makeidx}
\usepackage{amsfonts}
\usepackage{authblk} 
\usepackage{sectsty}
\usepackage{amsmath,amssymb,epsfig}
\usepackage{amscd}
\usepackage{amsthm}
\usepackage{mathrsfs}
\usepackage{dsfont}
\usepackage[applemac]{inputenc}
\usepackage[english]{babel}
\usepackage{enumitem} 
\usepackage[]{latexsym}
\usepackage{caption}
\usepackage{subfigure}
\usepackage{hyperref}
\usepackage{blindtext}
\usepackage{verbatim}
\usepackage{cancel}
\usepackage{cases}
\usepackage{cite}

\hypersetup{
	pdftitle={},%
	pdfauthor={},%
	pdfsubject={},%
	pdfkeywords={},%
	colorlinks=true,%
	linkcolor=blue,%
	citecolor=crimson,%
	linktocpage=true,%
	hyperfootnotes=true,%
	pageanchor=true
}

\definecolor{crimson}{rgb}{0.7, 0.08, 0.24}
\definecolor{blu}{rgb}{0.0, 0.18, 0.65}

\newcommand{\be}{\begin{equation}}

\newcommand{\ee}{\end{equation}}
\newcommand{\bes}{\begin{eqnarray}}
\newcommand{\ees}{\end{eqnarray}}
\def\bean{\begin{eqnarray*}}
	\def\eean{\end{eqnarray*}}
\def\nn{\nonumber}
\newcommand{\dd}{\mathrm{d}}

\newcommand{\p}{\partial}

\renewenvironment{thebibliography}[1]
{\section*{References}\frenchspacing\small
	\begin{list}{[\arabic{enumi}]}
		{\usecounter{enumi}\parsep=2pt\topsep 0pt
			\settowidth{\labelwidth}{[#1]}
			\leftmargin=\labelwidth\advance\leftmargin\labelsep
			\rightmargin=0pt\itemsep=1pt\sloppy}}{\end{list}}
\numberwithin{equation}{section}

\begin{document}
	\title{\textbf{\textsf{Quantum Corrected Polymer Black Hole Thermodynamics: Mass Relations and Logarithmic Entropy Correction}}}
	\author[1,3]{\textsf{Fabio M. Mele}\footnote{\texttt{fabio.mele@oist.jp}}}
	\author[2,3]{\textsf{Johannes M\"unch}\footnote{\texttt{johannes.muench@cpt.univ-mrs.fr}}}
	\author[3]{\textsf{Stratos Pateloudis}\footnote{\texttt{stratos.pateloudis@physik.uni-r.de}}\vspace{0.35cm}}  
	
	\affil[1]{\textsf{Okinawa Institute of Science and Technology, \protect\\1919-1 Tancha, Onna-son, Okinawa 904-0495, Japan.}\vspace{0.375cm}}
	
	\affil[2]{\textsf{Aix-Marseille Universit\'e, Universit\'e de Toulon, CNRS, CPT,
			\protect\\ 13288 Marseille, France.}\vspace{0.35cm}}
	
	\affil[3]{\textsf{Institute for Theoretical Physics, University of Regensburg, \protect\\ 93040 Regensburg, Germany.}\vspace{-0.5cm}}
			
	\date{\today}
	
	\maketitle
	
	\vspace{-0.85cm}
	\begin{abstract}
	In this paper, we continue the analysis of the effective model of quantum Schwarz\-schild black holes recently proposed by some of the authors in \cite{BodendorferMassandHorizon,Bodendorferbvtypevariables}. In the resulting quantum-corrected spacetime the central singularity is resolved by a black-to-white hole bounce, quantum effects become relevant at a unique mass-independent curvature scale, while they become negligible in the low curvature region near the horizon and classical Schwarzschild geometry is approached asymptotically. This is the case independently of the relation between the black and white hole masses, which are thus freely specifiable independent observables. A natural question then arises about the phenomenological implications of the resulting non-singular effective spacetime and whether some specific relation between the masses can be singled out from a phenomenological perspective. Here we focus on the thermodynamic properties of the effective polymer black hole and analyze the corresponding quantum corrections as functions of black and white hole masses. The study of the relevant thermodynamic quantities such as temperature, specific heat, and horizon entropy reveals that the effective spacetime generically admits an extremal minimal-sized configuration of quantum-gravitational nature characterized by vanishing temperature and entropy. For large masses, the classically expected results are recovered at leading order and quantum corrections are negligible, thus providing us with a further consistency check of the model. The explicit form of the corrections depends on the specific relationship among the masses. In particular, a first-order logarithmic correction to the black hole entropy is obtained for a quadratic mass relation. The latter corresponds to the case of proper finite-length effects which turn out to be compatible with a minimal length generalized uncertainty principle associated with an extremal Planck-sized black hole.
	\end{abstract}

		\section{Introduction}
		
		Black hole thermodynamics has a fascinating and long story by now \cite{HawkingBlackholesin, BekensteinBlackHolesandEntropy,BekensteinBlackholesand,HawkingParticlecreation}. 
		The seminal work goes back to Bekenstein who initiated this by claiming that black holes could potentially store information on their horizon surface. 
		This idea was based on the earlier derivation of Hawking's theorem according to which the area of a black hole horizon could never decrease \cite{HawkingBlackholesin}, leading in this way to a generalized notion of second law \cite{BekensteinGeneralizedsecondlaw}. 
		Furthermore was put into more solid foundations by the work of Hawking \cite{HawkingParticlecreation} together with Bardeen and Carter \cite{BardeenThefourlawsof}. For reviews see \cite{WaldBHthermodynamics,CarlipBHThermodynamics}. 
		However, the cost of setting the arena was the emergence of a paradox, the so-called black hole information paradox.  
		Since then, there is ongoing research trying to understand in more detail how these remarkable black objects acquire a thermodynamical description, what is the meaning of this paradox, and if eventually could be surpassed. It is widely believed that a full understanding of both the microscopic origin of black hole properties and the fate of the information-loss paradox would require a theory of quantum gravity. Over the years, insights and techniques from different approaches to quantum gravity have been used to attack the problem  \cite{MaldacenaEternalBH,GiddingsBHinformationunitaritynonlocality,AmadeiHawkingInformationPuzzle,AmadeiUnitarityandinformation,AshtekarInformationNotLost2DBH}, but no definite answer has been reached so far. For a review on this paradox independent of the approach see \cite{MathurTheInformationParadox}.  
		Even though these problems are fascinating by themselves, and different approaches lead to fruitful (or equivalent) realizations of the description we will not discuss them in full generality here. In particular, since a complete full theory of quantum gravity is still currently out of reach, to start investigating these questions in simplified models where quantum gravity effects can be efficiently incorporated becomes of great importance.
		
		In this work, we will focus on the thermodynamics of the so-called effective \textit{polymer} black holes. These are effective models of quantum corrected black hole spacetimes motivated by ideas and techniques borrowed from loop quantum gravity (LQG) \cite{RovelliQuantumGravity,RovelliBook2,ThiemannLecturesonLoop,ThiemannIntroductiontoModern,ThiemannBook,BodendorferAnElementaryIntroduction}.
		Specifically, the so-called \textit{polymerisation procedure} refers to a phase space regularisation capturing (some of) the relevant quantum corrections. Such a regularisation shares a similar spirit with holonomy corrections in lattice gauge theories according to which the field strength is approximated by holonomies of the gauge connection along the plaquettes that construct the lattice. 
		This approximation is based on the spirit that in the LQG description, one usually deals with a lattice which in addition is dynamical. The plaquettes, in this case, could be thought of as patches of dynamical geometries. Several models of effective polymer black holes have been developed in the last years \cite{AshtekarQuantumGeometryand,ModestoLoopquantumblack,CampigliaLoopquantizationof,ModestoSemiclassicalLoopQuantum,ModestoBlackHoleinterior,CorichiLoopquantizationof,BoehmerLoopquantumdynamics,ChiouPhenomenologicalloopquantum,OlmedoFromblackholesto,AshtekarQuantumTransfigurarationof,AshtekarQuantumExtensionof,Bouhmadi-LopezAsymptoticnon-flatness,BojowaldComment2,BodendorferAOSNote,BodendorferEffectiveQuantumExtended,Bodendorferbvtypevariables,BodendorferMassandHorizon,KellyEffectiveloopquantumgravity,KellyBlackholecollapse,FaraoniUnsettlingphysicsin,GanPropertiesofthe,BojowaldAno-goresult,GeillerSymmetriesofthe,SartiniQuantumdynamicsof}, most of which focus on the simplest case of a Schwarzschild black hole (see also \cite{CaravelliSpinningLBH,LiuShadowAndQuasinormal,BrahmaObservationalConsequences} for some results in the rotating setting). The model considered in this work is the effective quantum corrected Schwarzschild black hole introduced by some of the authors in \cite{BodendorferMassandHorizon,Bodendorferbvtypevariables}.
		The common feature of these types of models, in a nutshell, is that by implementing some regularization (exponentiated holonomies) on the variables that one is using to describe phase space functions a resolution of the black hole singularity is achieved, in a sense that the curvature is bounded across the whole spacetime, and at the same time the model remains effectively semiclassical. 
		This procedure achieves a twofold goal. 
		First of all, the rather classical singularity of conventional general relativity is replaced (since a theory of quantum gravity is expected to be ultraviolet-complete) by a transition surface that transfers an observer from the black-to-white-hole region repeatedly. The regions are called as such because one can prove that the relevant enclosing ``horizons'' are trapping and anti-trapping surfaces, respectively \cite{BodendorferMassandHorizon}. In this respect, however, it is worth mentioning that the polymerisation prescription is not unique. In particular, the simplest choice of a $\sin$ function commonly adopted in the literature based on e.g. holonomy arguments, curvature upper bounds, or preservation of certain classical symmetries (see e.g. \cite{AshtekarQuantumGeometryand,ModestoLoopquantumblack,CampigliaLoopquantizationof,ModestoSemiclassicalLoopQuantum,ModestoBlackHoleinterior,CorichiLoopquantizationof,BoehmerLoopquantumdynamics,ChiouPhenomenologicalloopquantum,OlmedoFromblackholesto,AshtekarQuantumTransfigurarationof,AshtekarQuantumExtensionof,Bouhmadi-LopezAsymptoticnon-flatness,BojowaldComment2,BodendorferAOSNote,BodendorferEffectiveQuantumExtended,Bodendorferbvtypevariables,BodendorferMassandHorizon,KellyEffectiveloopquantumgravity,KellyBlackholecollapse,FaraoniUnsettlingphysicsin,GanPropertiesofthe,BojowaldAno-goresult,GeillerSymmetriesofthe,SartiniQuantumdynamicsof}), captures only some of the quantum corrections. Indeed, as pointed out in \cite{Alesci:2018loi, Alesci:2019pbs, Alesci:2020zfi} in the context of quantum reduced loop gravity, taking the expectation value on coherent states peaked around classical spherically symmetric geometries of a partially gauge fixed Hamiltonian constraints on full LQG Hilbert space leads to an effective Hamiltonian constraint with a different polymerisation than the $\sin$-function usually involved in the analysis of effective polymer models, the latter being recovered only when some LQG quantum d.o.f. are neglected. Such different polymerisation prescriptions may result into a different spacetime structure where for instance no white hole horizon is formed.
		
		Second, beside of the specific details of the polymer model considered, the resulting description remains effective and the quantum corrected smooth spacetime geometry of such regular black holes can be described through an explicit metric. Hence, one can exploit the already well-developed machinery for semi-classical black holes and study several aspects of such effective geometries. One of these is thermodynamics, and in a satisfactory analogy with the conventional thermodynamics, a gravitational physicist expects that as the thermodynamic limit of a microscopic system described by statistical mechanics is realized, in the same way, the thermodynamic limit of a black hole emerges.
		The microscopic gravitational degrees of freedom in our case will remain unspecified since, at this effective level, no information is accessible whatsoever for the microscopic structure. Nevertheless, the main macroscopic consequences of the relevant quantum gravity effects can be captured at the effective level, and as such its study might unravel interesting phenomenological and observational implications\footnote{Some investigations for instance point towards a relation of primordial black hole remnants and dark matter (see e.g. \cite{BarrowThecosmologyof,CarrBlackholerelics,ModestoSelf-dualBlackHoles,RovelliSmallblackwhite} and references therein).}. In this respect, here we are interested in thermodynamics from an infrared approach because even semi-classically one can extract useful physics, and astrophysical measurements could potentially take place even at these scales \cite{LIGOObservationofGravitational,EHTFirstM87EHT}. 
		
		By several methods, and independent of the approach, one generically gets corrections on the semi-classical results of thermodynamic quantities. The two common routes are:
		a) either by a microscopic analysis using statistics suitable to analyze the degrees of freedom that count for the entropy, or b) by adding quantum corrections on, the already semi-classical, formulas for the four black hole thermodynamics laws.
		A common feature shared by the major quantum gravity approaches is the occurrence of logarithmic corrections next-to-leading-order in entropy, the leading order being provided by the classical Bekenstein-Hawking area law (for a sample from different approaches see e.g. \cite{CarlipLogarithmicCorretionsTo, MukherjiLogCorrectionAdSBHAdSCFT,KaulLogarithmicCorrectionto,StromingerMicroscopicOriginOf,StromingerBlackHoleEntropy,MannUniversalityofQuantum, RovelliBHentropyfromLQG,AshtekarQuantumGeometryAndBlackHoleEntropy,AshtekarQuantumgeometryof,KaulQuantumBHEntropy,BarberoQuantumisolatedhorizons,MeissnerBlackholeentropy,LivineQuantumblackholes,SenBHEntropyFunction,CaravelliHolographiceffectiveactions,JeonLogarithmiccorrectionsto} and references therein).
		The microscopic origin of this is due to a specific counting of the degeneracy of degrees of freedom that shape the black hole surface.  
		We shall see however that, without counting (macroscopically) degrees of freedom and being at an effective level, we might get a logarithmic term next to the leading order for the entropy also for our model by following method b).
		This is feasible because the model contains two \textit{independent} Dirac observables, respectively interpreted as the black and white hole masses for the two asymptotic Schwarzschild spacetimes smoothly connected by a transition surface replacing the classical singularity.
		The presence of such observables allows one to study how different choices of the relation between the two masses affect the quantum corrections to the thermodynamic quantities and whether for a particular choice it is possible to get logarithmic corrections.
		Such kind of question comes to be of twofold importance for the specific setting of the model considered here.
		Indeed, on the one hand, the recovery of the classical results at leading order for sufficiently large masses provides us with a further consistency check of the model by itself.
		On the other hand, it provides a way to select specific mass relations among the many possibilities based on their thermodynamic effects. 
		The study of the relevant thermodynamic quantities such as temperature, specific heat, and entropy reveals in fact that the classical results are approached for sufficiently large masses compared to the scales governing the onset of quantum effects, thus confirming that general relativity provides a good approximation at low curvatures.
		In the high curvature regime, instead, quantum effects induce significant departures from the classical behaviors.
		The upshot is that the polymer black hole settles into an extremal static object of Planck's area as we are lowering its mass.
		At exactly this point, its horizon area coincides with the cross-section of the transition surface with vanishing temperature and entropy.
		This is the case independently of the specific choice of a relationship between the black and white hole masses.
		The specific form of the next-to-leading-order corrections of the large mass expansion depends instead on the mass relation.
		In particular, a quadratic mass relation $M_{WH}\propto M_{BH}^2$ yields a logarithmic correction to the entropy. 
		Interestingly, such a mass relation corresponds to the case in which a strict distinction between large curvatures and small-length quantum effects can be made. 
		In this sense, such a mass relation comes to be thermodynamically special.
		
		The paper is organized as follows. In Sec. \ref{Modelreview} we recall the main features of the effective polymer Schwarzschild black hole presented in \cite{BodendorferMassandHorizon,Bodendorferbvtypevariables}. The thermodynamic properties of such a black hole are studied in Sec. \ref{Sec:BHtherm} which constitutes the main part of the manuscript.
		The temperature is derived and discussed in Sec.~\ref{Sec:quantcorrectedtemperature}, followed by the derivation of the specific heat in Sec.~\ref{sec:specificheat}.
		The entropy and its dependence on the mass relation are analyzed in Sec.~\ref{sec:entropy}. The thermodynamic investigation closes with the derivation of the evaporation time in Sec.~\ref{sec:evaporationtime}.
		We close in Sec. \ref{Sec:conclusion} with some concluding remarks and discussing future directions. Some supplementary calculations and the discussion of the near horizon geometry are reported in Appendix \ref{App:nearhorizon&Euclidean}.
		
		\section{The Model}\label{Modelreview}
		
		Let us first start by briefly recalling the main ideas and results of the effective model for polymer Schwarzschild black holes developed in \cite{BodendorferMassandHorizon,Bodendorferbvtypevariables}.
		We will mainly focus on those aspects of the model, which are relevant for the present analysis, thus referring to \cite{BodendorferMassandHorizon,Bodendorferbvtypevariables} for a detailed exposition.
		\subsection{Classical Theory}\label{Sec:classicaltheory}
		The starting point is a metric of the form \cite{CavagliaHamiltonianFormalismFor,VakiliClassicalPolymerizationof}
		\begin{equation}\label{clmetric2}
		\dd s^2=-\bar a(r)\dd t^2+N(r)\dd r^2+2\bar B(r)\dd t\dd r+\bar b^2(r)\dd\Omega_2^2\;, 
		\end{equation}
		which describes a generic static and spherically symmetric 4-dimensional spacetime of topology $\mathbb R\times\mathbb R\times \mathbb S^2$, and $\dd\Omega_2^2 = \dd\theta^2+\sin^2\theta\dd\phi^2$ being the round metric on the $r = const.$, $t = const.$ 2-sphere.
		Inserting this ansatz into the Einstein-Hilbert-action yields, after neglecting boundary terms, the first order Lagrangian ($G=c=1$)
		\begin{equation}\label{BHlagrangian1}
		L(\bar a,\bar b,\bar n)=2L_o\sqrt{\bar n}\left(\frac{\bar a'\bar b'\bar b}{\bar n}+\frac{\bar a\bar b'^2}{\bar n}+1\right)\;,
		\end{equation}
		where primes denote derivatives w.r.t. $r$, $\bar n$ is a Lagrange multiplier which reflects the gauge freedom in the definition of the coordinates $r$ and $t$ and it is related to the lapse $N$ w.r.t. the foliation in $r$-slices via the relation $\bar n = \bar a N + \bar B^2$. The otherwise-divergent integral in the non-compact $t$-direction has been regularised by introducing a fiducial cell in the constant $r$ slices of topology $[0,L_o]\times\mathbb S^2$, $L_o$ playing the role of an infrared cutoff. By smearing the dynamical variables in $t$-direction, that is by introducing the integrated variables
		\be
		\sqrt{n}= \int_{0}^{L_o} \dd t \sqrt{\bar n} =L_o\sqrt{\bar n}\quad,\quad\sqrt{a} = \int_{0}^{L_o} \dd t \sqrt{\bar a} =L_o\sqrt{\bar a}\quad,\quad b=\bar b\quad,\quad B= \int_{0}^{L_o} \dd t \bar{B} =L_o\,\bar B\;,
		\ee
		$L_o$-factors in the Lagrangian \eqref{BHlagrangian1} can be absorbed thus yielding
		\begin{equation}\label{BHlagrangian2}
		L(a,b,n)=2\sqrt{n}\left(\frac{a'b'b}{n}+\frac{a b'^2}{n}+1\right)\;.
		\end{equation}
		In what follows, we will distinguish between the coordinate length $L_o$ of the fiducial cell in $t$-direction, which as such depends on the choice of the chart, and the coordinate-independent physical length $\mathscr L_o$ of the fiducial cell at a certain reference point $r_{\text{ref}}$ given by $\mathscr L_o := \left.\sqrt{a}\right|_{r=r_{\text{ref}}}=L_o\left.\sqrt{\bar a}\right|_{r=r_{\text{ref}}}$. Both $L_o$ and $\mathscr L_o$ are fiducial structures and of course physical quantities must not depend on them.
		After performing a Legendre-transformation, the canonical analysis of the system can be carried out employing the standard steps of the constraint algorithm for reparametrization-invariant systems thus leading to a 4-dimensional phase space spanned by the configuration variables $a$,$b$ and their canonically conjugate momenta $p_a$, $p_b$ with non-vanishing Poisson brackets $\{a,p_a\}=1=\{b,p_b\}$ and subject to the Hamiltonian (first-class) constraint 
		\begin{equation}\label{hamiltonian1}
		H_{cl}=\sqrt{n} \mathcal{H}_{cl} \qquad, \qquad \mathcal{H}_{cl} = \frac{p_ap_b}{2b}-\frac{ap_a^2}{2b^2}-2 \approx 0\;,
		\end{equation}
		which reflects the fact that $n$ appears as Lagrange multiplier encoding the freedom in changing the $r$ coordinate, i.e. that $N$ and $\bar B$ in \eqref{clmetric2} are pure gauge degrees of freedom.  
		
		Crucial in the construction of our model is the  canonical transformation to the following phase space variables
		\begin{align}
		v_k=\frac{2b^4}{p_a}\qquad&,\qquad k=\frac{a' b'}{n\,b}=\frac{p_a}{2b^3}\left(\frac{p_b}{2}-\frac{ap_a}{b}\right)\;,\label{defvariablesksector}\\
		v_j=\frac{2 b^2}{p_a}\left(2ap_a - \frac{b p_b}{2}\right)\qquad&,\qquad j=\frac{b'}{\sqrt{n}\,b}=\frac{p_a}{2b^2}\;,\label{defvariablesjsector}
		\end{align}
		in terms of which the Hamiltonian constraint \eqref{hamiltonian1} takes the remarkably simple form
		\begin{equation}\label{hamiltonian2}
		H_{cl}=\sqrt{n}\mathcal{H}_{cl} \qquad,\qquad \mathcal{H}_{cl} = 3v_k kj +v_jj^2-2 \approx 0\;.
		\end{equation}
		Solving the corresponding equations of motion and plugging the result into the expression of the metric coefficients obtained by inverting the definitions \eqref{defvariablesksector}, \eqref{defvariablesjsector}, it is easy to show that the standard form of Schwarzschild metric is recovered for $N(b)=(1-\frac{2M}{b})^{-1}$, i.e. $B^2=Na-n=0$, where $M$ is the black hole ADM mass. We refer to \cite{BodendorferMassandHorizon} for details and explicit solutions.
		Note that under rescaling of the fiducial length $L_o \mapsto \alpha L_o$, i.e. $\mathscr L_o \mapsto \alpha \mathscr L_o $, the variables \eqref{defvariablesksector}, \eqref{defvariablesjsector} change accordingly as $v_k\mapsto \alpha\,v_k$, $v_j\mapsto\alpha^2\,v_j$, $k\mapsto\,k$, $j\mapsto \alpha^{-1}\,j$. Of course, physical quantities in the end must be independent of any fiducial cell rescaling.
		The main advantage of these new variables relies on the physical interpretation for the canonical momenta. Indeed, the fiducial cell-independent on-shell quantities
		\begin{equation}\label{Pinterp}
		k(b)=\frac{2M}{b^3} \qquad,\qquad j(b) \mathscr L_o=\frac{1}{b} \left(\frac{D}{\mathscr L_o}\right)^\frac{1}{2} \;,
		\end{equation}
		are respectively proportional to the square root of the on-shell value of the Kretschmann scalar $\mathcal K=48M^2/b^6$ and the inverse areal radius, the latter being related to the angular components of the extrinsic curvature as $1/b=\sqrt{N(b)}K_\phi^\phi=\sqrt{N(b)}K_\theta^\theta$.
		Here $D$ is an integration constant that remains unspecified in the classical setting. Note that the above on-shell interpretation for the $j$-sector requires $D$ to be mass independent\footnote{While the classical physical description is not affected by the specific choice of $D$ which can be washed away of the line element by a coordinate redefinition \cite{BodendorferMassandHorizon}, in the effective quantum theory it will influence the onset of quantum effects.}.
		From an off-shell point of view, the variable $k$ can be related to the Riemann tensor via the relation
		\begin{equation}\label{eq:kMMS}
		k \stackrel{\mathcal{H}\approx 0}{\approx} R_{\mu \nu \alpha \beta} \epsilon^{\mu \nu} \epsilon^{\alpha \beta} = b \left(1- \frac{b'^2}{N}\right) = \frac{2 M_{\text{Misner-Sharp}}(b)}{b^3}\;,
		\end{equation}
		\noindent
		where $\epsilon^{\mu \nu} = g^{\mu \alpha} g^{\nu \beta} \epsilon_{\alpha \beta}$ with $\epsilon_{\alpha \beta} \dd x^\alpha \wedge \dd x^{\beta} = b^2 \sin\theta \dd \theta \wedge \dd \phi$ is the volume two-form of the $r,t=const.$ two-sphere and $M_{\text{Misner-Sharp}}$ is the Misner-Sharp mass (see e.g. \cite{SzabadosQuasi-LocalEnergy} and references therein). $M_{\text{Misner-Sharp}}(b)$ measures the gravitational mass enclosed in the constant $t$-sphere of areal radius $b$.
		Thus, the definitions \eqref{defvariablesksector}, \eqref{defvariablesjsector} together with the on-shell relations \eqref{Pinterp} provide us with a set of canonical variables whose canonical momenta are directly related to the Kretschmann scalar and the extrinsic curvature, respectively. As discussed in detail in \cite{BodendorferMassandHorizon} and will be reviewed in Sec.~\ref{sec:effectiveQT}, a polymerisation scheme based on these variables turns out to be well suited for achieving a unique curvature upper bound at which quantum effects become dominant.
		
		\subsection{Effective Quantum Theory}\label{sec:effectiveQT}
		Using the fact that $r$ is a time-like coordinate in the black hole interior ($a < 0, N < 0$), the latter can be foliated by homogeneous space-like three-dimensional $\mathbb R\times\mathbb S^2$ hypersurfaces and it is isometric to the vacuum Kantowski-Sachs cosmological spacetime. 
		Techniques from homogeneous and anisotropic LQC can be thus imported to construct an effective quantum theory of the minisuperspace under consideration.
		In the spirit of LQC, the semi-classical features of the model should be captured by an effective Hamiltonian obtained by polymerizing the canonical momenta which amount to replace them by functions of their point holonomies. A commonly adopted and particularly simple polymerisation choice in LQC-literature (see e.g. \cite{AshtekarMathematicalstructureof,AshtekarLoopquantumcosmology:astatusreport} and references therein) is the sin function. According to the above on-shell interpretation of the momenta \eqref{Pinterp}, a polymerisation $k \mapsto \sin\left(\lambda_k\, k\right)/\lambda_k$, $j \mapsto \sin\left(\lambda_j\, j\right)/\lambda_j$ with (constant) polymerisation scales $\lambda_k$, $\lambda_j$  regularises respectively the large curvature and small areal radius regimes ($\lambda_k k,\lambda_j j\sim1$), while the classical behaviour is recovered at low curvatures and large distances ($\lambda_k k,\lambda_j j\ll1$).
		Note that while $\lambda_k$, $\lambda_j$  transform under fiducial cell rescaling, the quantities $\lambda_k$ and $\lambda_j / \mathscr L_o$ are coordinate- and fiducial cell-independent and are considered of order inverse Planck-curvature and Planck-length, respectively.
		Choosing the Lagrange multiplier $n = const. = \mathscr L_o^2$ and $B = 0$, the polymerised  Hamiltonian system is analytically solvable.  The resulting quantum corrected effective line element is given by \cite{BodendorferMassandHorizon}
		\be\label{metric}
		\dd s^2=-\frac{a(x)}{\lambda_j^2}d t^2+\frac{\lambda_j^2}{a(x)}d x^2+b(x)^2d\Omega^2_2,
		\ee
		\noindent
		where we rescaled the $t$- and $r$-coordinate according to $t \mapsto \lambda_j t/L_o $ and $x = \mathscr{L}_o r/\lambda_j$, and the metric coefficients read as
		\allowdisplaybreaks
		\begin{align}
		b^2(x)&=v_k(r)\frac{\sin(\lambda_jj)}{\lambda_j},
		\notag
		\\
		&=\frac{1}{2} \left(\frac{\lambda_k}{M_{BH} M_{WH}}\right)^{\frac{2}{3}}\frac{1}{\sqrt{1+x^2}}\frac{M_{BH}^2\left(x+\sqrt{1+x^2}\right)^6+M_{WH}^2}{\left(x+\sqrt{1+x^2}\right)^3}\;, \label{eq:bsol}
		\\
		a(x)&=\frac{1}{2v_k}\frac{\lambda_j^2}{\sin^2(\lambda_jj)}\left(v_j\frac{\sin(\lambda_jj)}{\lambda_j}+v_k\frac{\sin(\lambda_kk)}{\lambda_k}\right),
		\notag
		\\
		&=2 \lambda_j^2 \left(\frac{M_{BH} M_{WH}}{\lambda_k}\right)^{\frac{2}{3}}\left(1-\left(\frac{M_{BH} M_{WH}}{\lambda_k}\right)^{\frac{1}{3}}\frac{1}{\sqrt{1+x^2}}\right)\frac{\left(1+x^2\right)^{\frac{3}{2}}\left(x+\sqrt{1+x^2}\right)^3}{M_{BH}^2\left(x+\sqrt{1+x^2}\right)^6+M_{WH}^2}\,,\label{eq:asol}
		\\
		k(x)&=\frac{2}{\lambda_k}\cot^{-1}\left(\frac{M_{BH}}{M_{WH}}\left(x+\sqrt{1+x^2}\right)^3\right)\;,\label{effdynsol3}
		\\
		j(x)&=\frac{1}{\lambda_j}\cot^{-1}\left(x\right)+\frac{\pi}{\lambda_j}\theta\left(-x\right)\;,\label{effdynsol4}
		\end{align}
		\noindent
		where the solution for $v_k$ and $v_j$ are not explicitly reported here and $M_{BH}$ and $M_{WH}$ are the two integration constants in the system.
		The names become meaningful when studying the asymptotic spacetimes (the coordinate $x$ has range $\left(-\infty,\infty\right)$) as briefly discussed below.
		Before moving at this, few remarks are order.
		
		First of all, let us notice that, after inserting the expressions \eqref{eq:bsol}, \eqref{eq:asol} into the line element \eqref{metric}, the final metric does not depend on the polymerisation scale $\lambda_j$, the latter just entering the rescaling of $t$- and $r$-coordinates.
		Therefore, no physical results will be dependent on $\lambda_j$.
		
		Second, it is worth noticing that the above line element is well-defined and smooth for the whole $x$-range, that is for both the interior and exterior regions.
		This is due to the fact that the starting point for the Hamiltonian formulation discussed in Sec. \ref{Sec:classicaltheory} was based on inserting the stationary ansatz \eqref{clmetric2} for the metric into the Einstein-Hilbert action, leading to an effective point particle Lagrangian with derivatives in $x$-direction (or equivalently $r$-direction). The vector field $\partial_x$ is space-like outside of the black hole horizon and time-like in the interior region.
		However, since the symmetry-reduced system under consideration is formally equivalent to a point particle system, it is possible to construct a Hamiltonian formulation that is valid everywhere, outside and inside.
		The resulting Hamiltonian generates ``evolution'' along the $x$-direction, with the corresponding Hamiltonian flow being space-like and time-like in the exterior and interior region, respectively.
		Such a Hamiltonian is thus equivalent to the ADM-Hamiltonian for the interior but has no ADM-counterpart in the exterior.
		However, this is still a valid formalism as classically the Schwarzschild metric is recovered for both regions.
		Another way to view this is to solve the system only for the interior region with a well-defined notion of time-like Hamiltonian flow and then smoothly extend the solution to the exterior. This leads to the same solutions of the present Hamiltonian formulation.
		
		In the effective theory, the polymerisation procedure is performed (at the off-shell level) in the interior, where a well-defined time-like Hamiltonian vector field is available and, once the equations of motion are solved, the resulting effective line element is smoothly extended to the exterior.
		Alternatively, we could ask for a modification of the equations of motion compatible with the polymerisation in the interior.
		This would give exactly the same result as in Eqs. \eqref{eq:bsol}-\eqref{effdynsol4}.
		Therefore, in the present Hamiltonian formulation, there is no need to make any formal distinction between the interior and exterior regions, the on-shell results being smooth and well defined over the whole $x$-domain. As we will discuss in the remaining of this section, the metric is well-defined globally and can be viewed as a regular eternal black hole from a global perspective. In particular, all relevant notions for the later purposes of thermodynamics such as ADM-mass, etc. are well-defined as the metric is asymptotically flat. This is in fact also suggested by the existence of Dirac observables whose on-shell interpretation is the mass of the asymptotic Schwarzschild geometry.
		
		To see this, let us recall then the structure of the spacetime described by the metric \eqref{metric}-\eqref{eq:asol} starting with the asymptotic spacetimes at $x \rightarrow \pm \infty$. Expressing $a$ in terms of the areal radius $b$ in the limit $x \rightarrow +\infty$ yields
		\begin{equation}
		a\left(b\left(x \rightarrow +\infty\right)\right) = \lambda_j^2 \left(\frac{M_{WH}}{8 \lambda_k M_{BH}^2}\right)^{\frac{2}{3}}\left(1- \frac{2 M_{BH}}{b_+}\right) \quad , \quad b_+^2 := b^2\left(x \rightarrow +\infty\right) = \left(8 \lambda_k \frac{M_{BH}^2}{M_{WH}}\right)^{\frac{2}{3}} x^2\,,
		\end{equation}
		\noindent
		and therefore to the asymptotic line element
		\begin{equation}\label{eq:asympotic+}
		\dd s^2 \simeq - \left(\frac{M_{WH}}{8 \lambda_k M_{BH}^2}\right)^{\frac{2}{3}}\left(1- \frac{2 M_{BH}}{b_+}\right)
		\dd t^2 + \left(1-\frac{2 M_{BH}}{b_+}\right)^{-1}\dd b_+^2+ b_+^2 \dd \Omega^2_2 \;.
		\end{equation}
		\noindent
		Rescaling in a last step $t \mapsto \tau = \left(\frac{M_{WH}}{8 \lambda_k M_{BH}^2}\right)^{\frac{1}{3}} t$ gives asymptotically the Schwarzschild metric with black hole mass $M_{BH}$, namely
		\begin{equation}
		\dd s^2 \simeq - \left(1- \frac{2 M_{BH}}{b_+}\right)
		\dd \tau^2 + \left(1-\frac{2 M_{BH}}{b_+}\right)^{-1}\dd b_+^2 + b_+^2 \dd \Omega^2_2 \;,
		\end{equation}
		\noindent
		which justifies the notation $M_{BH}$ for this integration constant and shows that $\tau$ is the Schwarzschild time on the black hole side $x \rightarrow +\infty$.
		The same analysis is possible in the $x \rightarrow -\infty$ limit, leading to the asymptotic metric coefficients
		\begin{equation}
		a\left(b\left(x \rightarrow -\infty\right)\right) = \lambda_j^2 \left(\frac{M_{BH}}{8 \lambda_k M_{WH}^2}\right)^{\frac{2}{3}}\left(1- \frac{2 M_{WH}}{b_-}\right) \quad , \quad b_-^2 := b^2\left(x \rightarrow -\infty\right) = \left(8 \lambda_k \frac{M_{WH}^2}{M_{BH}}\right)^{\frac{2}{3}} x^2\,,
		\end{equation}
		\noindent
		and therefore to the asymptotic spacetime
		\begin{equation}
		\dd s^2 \simeq - \left(\frac{M_{BH}}{8 \lambda_k M_{WH}^2}\right)^{\frac{2}{3}}\left(1- \frac{2 M_{WH}}{b_-}\right)
		\dd t^2 + \left(1-\frac{2 M_{WH}}{b_-}\right)^{-1}\dd b_-^2 + b_-^2 \dd \Omega^2 _2\;.
		\end{equation}
		\noindent
		Again, rescaling the time coordinate $t \mapsto \tau = \left(\frac{M_{BH}}{8 \lambda_k M_{WH}^2}\right)^{\frac{1}{3}} t$ gives asymptotically the Schwarzschild metric with mass $M_{WH}$
		\begin{equation}
		\dd s^2 \simeq - \left(1- \frac{2 M_{WH}}{b_-}\right)
		\dd \tau^2 + \left(1-\frac{2 M_{WH}}{b_-}\right)^{-1}\dd b_-^2  + b_-^2 \dd \Omega^2_2 \;.
		\end{equation}
		\noindent
		As this asymptotic region lies behind a transition from a trapped black hole interior to an anti-trapped interior region (cfr. Eq. \eqref{expansions} below), this is called the white hole side, which is classically not present. To see this, let us continue the analysis of the causal structure of the full metric \eqref{metric}. As can easily be deduced from Eq.~\eqref{eq:asol}, there exist two Killing horizons at which 
		\begin{equation}\label{eq:xhorizon}
		a(x_\pm) = 0 \qquad \text{for} \qquad x_\pm = \pm \sqrt{\left(\frac{M_{BH} M_{WH}}{\lambda_k}\right)^{\frac{2}{3}}-1} \,.
		\end{equation}
		\noindent
		Note that the horizons exist only if 
		\begin{equation}
		M_{BH} M_{WH} \ge \lambda_k\;,
		\end{equation}
		\noindent
		and an extremal configuration in which both horizons coincide is reached in case of the equal sign.
		The areal radius of the horizons is given by $b(x_\pm)$, which can be written as
		\begin{align}\label{eq:bhorizon}
		b\left(r_s^{(\pm)}\right)^2 =&\; \frac{M_{BH}}{2 M_{WH}} \left( \left(M_{BH} M_{WH}\right)^{\frac{1}{3}} \pm \sqrt{ \left(M_{BH} M_{WH}\right)^{\frac{2}{3}} - \lambda_k^{\frac{2}{3}}} \right)^3
		\notag
		\\
		&\;+ \frac{M_{WH}}{2 M_{BH}}  \frac{\lambda_k^2}{\left( \left(M_{BH} M_{WH}\right)^{\frac{1}{3}} \pm \sqrt{ \left(M_{BH} M_{WH}\right)^{\frac{2}{3}} - \lambda_k^{\frac{2}{3}}} \right)^3} \;,
		\end{align}
		\noindent
		and, for $M_{BH} M_{WH} \gg \lambda_k$, reduces to
		\begin{align}
		b(r_s^{(+)})^2 &\simeq 4 M_{BH}^2 \left[ 1 - \frac{3}{4} \left(\frac{\lambda_k}{M_{BH} M_{WH}}\right)^\frac{2}{3} - \mathcal{O}\left(\frac{\lambda_k^2}{M_{BH}^2 M_{WH}^2}\right)\right] \;, \label{eq:brspapprox}
		\\
		b(r_s^{(-)})^2 &\simeq 4 M_{WH}^2 \left[ 1 - \frac{3}{4} \left(\frac{\lambda_k}{M_{BH} M_{WH}}\right)^\frac{2}{3} - \mathcal{O}\left(\frac{\lambda_k^2}{M_{BH}^2 M_{WH}^2}\right)\right] \;, \label{eq:brsmapprox}
		\end{align}
		\noindent
		in agreement with the classical result on both sides at leading order and
		sub-leading (negative) quantum corrections suppressed by $\lambda_k$.
		Hence, for sufficiently large masses, the quantum corrected horizon is sightly smaller than the classical expectation.
		Another important surface in the effective spacetime is the so-called transition surface.
		This is defined as the space-like hypersurface $\mathcal T$ identified by the minimal reachable areal radius $b_{\mathcal{T}} = b(x_{\mathcal{T}})$ with $x_{\mathcal T}$ denoting the solution of the equation
		\begin{equation}
		\left.\frac{\dd b}{\dd x}\right|_{x = x_{\mathcal{T}}} = 0 \;.
		\end{equation}
		The resulting expressions are involved but can be solved analytically. The above condition in fact can be manipulated to arrive at a fourth-order equation, which has always exactly one real solution in the relevant domain. The explicit form of the solution is not needed in what follows and we refer then the interested reader to \cite{BodendorferMassandHorizon} for details.
		The name transition surface comes from the fact that the effective spacetime undergoes a transition from a black hole (BH) interior region, which is trapped, to a white hole (WH) interior region, which is anti-trapped, as it can be seen by computing the in- and out-going expansions $\theta^\pm$ given by
		\begin{equation}\label{expansions}
		\theta^\pm= - \sqrt{-\frac{2 a(x)}{\lambda_j^2}} \frac{b'(x)}{b(x)} \;,
		\end{equation}
		\noindent
		which are both negative for $x > x_{\mathcal{T}}$ and positive for $x < x_{\mathcal T}$.
		\begin{figure}[t!]
			\centering 
			\includegraphics[scale=0.6]{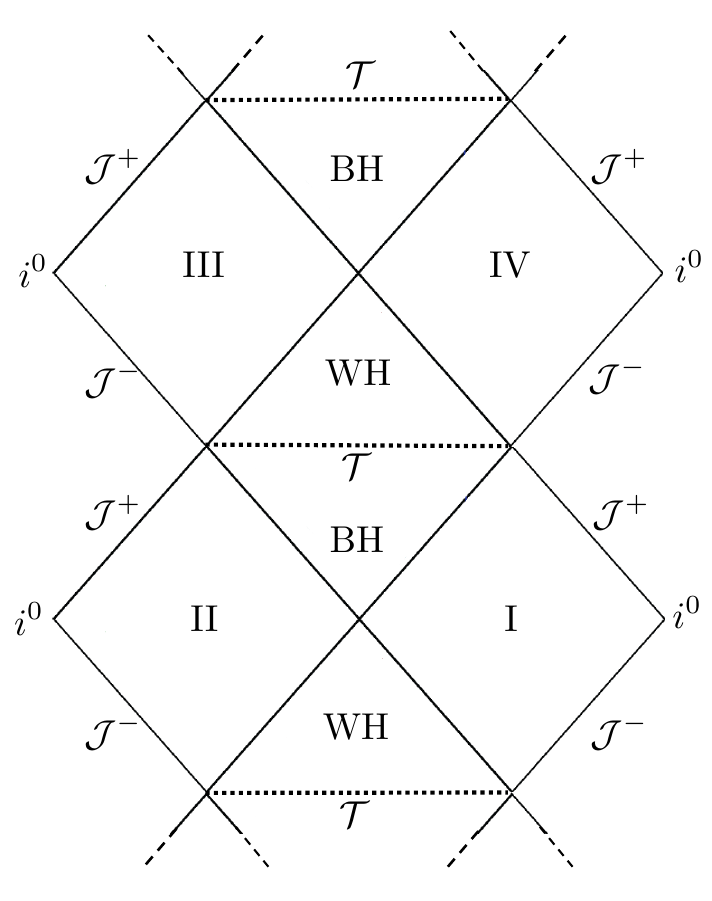}
			\caption{Penrose diagram for the Kruskal extension of the effective quantum corrected Schwarzschild black hole.}
			\label{penrosediagram}
		\end{figure}
		In the effective quantum corrected spacetime, the classical singularity is thus replaced by a transition surface interpolating between two asymptotic Schwarzschild geometries with mass $M_{BH}$ and $M_{WH}$, respectively. The causal structure of this spacetime can be summarised into the Penrose diagram reported in Fig. \ref{penrosediagram} obtained using generalized Kruskal-Szekeres coordinates. The details of the construction can be found in \cite{BodendorferMassandHorizon,BodendorferEffectiveQuantumExtended} and are not needed here.
		Let us just notice that this diagram consists of an infinite tower of ``Schwarzschild-like'' diagrams with infinitely many transitions between black and white hole interior regions.
		An important feature of this model is the fact that there are two \textit{independent} mass integration constants $M_{BH}$ and $M_{WH}$.
		These are not fixed by the canonical analysis or the resulting effective dynamics and therefore every relation between them is in principle allowed. From a phase space perspective, this reflects into the existence of the following two independent Dirac observables 
		\begin{align}
		2\mathcal{M}_{BH}&=\frac{\sin(\lambda_kk)}{\lambda_k}\cos\left(\frac{\lambda_kk}{2}\right)\left(\frac{2v_k}{\lambda_j\cot\left(\frac{\lambda_jj}{2}\right)}\right)^{\frac{3}{2}}\;,\\
		2\mathcal{M}_{WH}&=\frac{\sin(\lambda_kk)}{\lambda_k}\sin\left(\frac{\lambda_kk}{2}\right)\left(\frac{2v_k}{\lambda_j}\cot\left(\frac{\lambda_jj}{2}\right)\right)^{\frac{3}{2}}\;,
		\end{align}
		\noindent
		which as such commute with the Hamiltonian constraint $\left\{\mathcal M_{BH}, \mathcal H\right\} = \left\{\mathcal M_{WH}, \mathcal H\right\} = 0$, and reduce on-shell to the values of the integration constants, i.e.
		\be
		\mathcal M_{BH} \stackrel{\text{on-shell}}{=} M_{BH} \qquad , \qquad \mathcal M_{WH} \stackrel{\text{on-shell}}{=} M_{WH} \;.
		\ee
		\noindent
		As a consequence of the polymerisation with $\sin$-functions, the model is symmetric under mass exchange off- and on-shell.
		At the level of solutions, this can be seen by noticing that the metric \eqref{metric}-\eqref{eq:asol} is invariant under the transformation
		\begin{equation}
		M_{BH} \longmapsto M_{WH} \qquad , \qquad M_{WH} \longmapsto M_{BH} \qquad , \qquad x \longmapsto -x \;.
		\end{equation}
		\noindent
		More details are discussed in \cite{BodendorferMassandHorizon}.
		Although there is no canonical way to fix a relation between the two masses, two particular relations are special.
		To see this, let us notice that the classical regime is reached when
		\be\label{clasregimeBH}
		x\gg1\qquad,\qquad 8\,\frac{M_{BH}}{M_{WH}}\, x^3\gg1 \;,
		\ee
		\noindent
		on the black hole side and equivalently
		\be\label{clasregimeWH}
		|x|\gg1\qquad,\qquad 8\, \frac{M_{WH}}{M_{BH}}\, |x|^3\gg1\;,
		\ee
		\noindent
		on the white hole side.
		Expressed in terms of $b_+$ and $b_-$ this gives
		\begin{equation}\label{eq:classregimeb+}
		b_+ \gg \left(8 \lambda_k \frac{M_{BH}^2}{M_{WH}}\right)^\frac{1}{3} \qquad , \qquad \mathcal K_{cl}^{BH} = \frac{48 M_{BH}^2}{b_+^6} \ll \frac{48}{\lambda_k^2} \;,
		\end{equation}
		\noindent
		and 
		\begin{equation}\label{eq:classregimeb-}
		b_- \gg \left( 8 \lambda_k \frac{M_{BH}^2}{M_{WH}} \right)^{\frac{1}{3}} \frac{M_{WH}}{M_{BH}} = \left( 8 \lambda_k \frac{M_{WH}^2}{M_{BH}} \right)^{\frac{1}{3}}
		\quad , \quad \mathcal K_{cl}^{WH} = \frac{48 M_{WH}^2}{b_-^6} \ll \frac{48}{\lambda_k^2} \;.
		\end{equation}
		\noindent
		From this it becomes clear -- and this is the main achievement of this model -- that large curvature quantum effects become always relevant at the unique curvature scale $48/\lambda_k^2$ for any mass relation.
		Moreover, this scale is the same for both sides.
		Nevertheless, the first conditions in Eqs.~\eqref{eq:classregimeb+} and \eqref{eq:classregimeb-} might spoil the above symmetry in the onset of quantum effects by making the $j$-sector small $b$ effects more or less dominant on one side or the other depending on the asymmetry of the two masses.
		This singles out the following two special scenarios of preferred mass relations:
		\begin{itemize}
			\item The first option is a symmetric relation, i.e.
			\begin{equation}\label{symmassrelation}
			M_{BH} = M_{WH} \,,
			\end{equation}
			\noindent
			in which case both sides are exactly equivalent.
			The first conditions in Eqs.~\eqref{eq:classregimeb+} and \eqref{eq:classregimeb-} in fact become 
			\begin{align}
			&b_+ \gg \left(8 \lambda_k M_{BH}\right)^\frac{1}{3} \qquad \Rightarrow \qquad K_{cl}^{BH} = \frac{48 M_{BH}^2}{b_+^6} \ll \frac{3}{4 \lambda_k^2} \;,\\
			&b_- \gg \left(8 \lambda_k M_{WH}\right)^\frac{1}{3} \qquad \Rightarrow \qquad K_{cl}^{WH} = \frac{48 M_{WH}^2}{b_+^6} \ll \frac{3}{4 \lambda_k^2} \;,
			\end{align}
			\noindent
			which is again a unique curvature scale on both sides.
			Therefore, for $M_{BH}=M_{WH}$, both $j$- and $k$-polymerisations yield large curvature quantum effects respectively driven by the curvature scales $3/4\lambda_k^2$ and $48/\lambda_k^2$.
			\item Another possibility is a quadratic mass relation
			\begin{equation}\label{eq:quadraticrelation}
			M_{WH} = \frac{M_{BH}^2}{m} \;,
			\end{equation}
			\noindent
			and the reversed relation 
			\begin{equation}
			M_{BH} = \frac{M_{WH}^2}{m} \;,
			\end{equation}
			\noindent
			where $m$ is a constant of dimension mass. Let us focus on the relation Eq.~\eqref{eq:quadraticrelation} as the analysis is equivalent for both.
			In this case, the quantum effects on the BH side become relevant when
			\begin{equation}\label{eq:classregimeb+quadratic}
			b_+ \sim \left(8 \lambda_k m\right)^\frac{1}{3} \qquad , \qquad \mathcal K_{cl}^{BH} = \frac{48 M_{BH}^2}{b_+^6} \sim \frac{48}{\lambda_k^2} \;,
			\end{equation}
			\noindent
			and are thus driven by the unique curvature scale $48/\lambda_k^2$ and the critical length scale $\ell_{crit} = \left(8 \lambda_k m\right)^\frac{1}{3}$.
			This is also the case in which $j(b(x\rightarrow \infty)) \simeq 1/b$ with purely constant mass-independent proportionality factors and the canonical momentum $j$ can be truly interpreted as an inverse length or as the angular component of the extrinsic curvature.
			On the white hole side, the conditions Eq.~\eqref{eq:classregimeb-} become instead
			\be
			b_- \gg \left( 8 \lambda_k m \right)^{\frac{1}{3}} \frac{M_{WH}}{M_{BH}} = \ell_{crit} \frac{M_{WH}}{M_{BH}}
			\qquad , \qquad \mathcal K_{cl}^{WH} = \frac{48 M_{WH}^2}{b_-^6} \ll \frac{48}{\lambda_k^2} \;,
			\ee 
			\noindent
			i.e., the critical length scale $\ell_{crit}$ is shifted by the mass-asymmetry, while the  curvature scale remains unique.
			
		\end{itemize}	
		
		In the following the thermodynamic properties of this effective quantum black hole solution are analyzed for generic mass relations $M_{WH} = M_{WH}\left(M_{BH}\right)$, which gives another semi-classical perspective on these relations.
		A special focus lies on the analysis of how thermodynamic quantities as temperature, specific heat, and entropy depend on this relation.
		
		\section{Quantum Corrected Black Hole Thermodynamics}\label{Sec:BHtherm}
		
		In this section, we perform the thermodynamic analysis of the effective quantum corrected models for stationary spherically symmetric Schwarzschild black hole discussed in Sec.~\ref{sec:effectiveQT}.
		As the model does in principle not fix any relation between the two masses, our main interest relies on understanding whether certain relations between the black and white hole masses can be selected by studying the thermodynamic properties of this effective geometry.
		This can then further be related to the two special relations favored by studying the onset of quantum effects.
		Therefore, in the following, we assume that $M_{WH}$ is given always in terms of a relation $M_{WH}(M_{BH})$.
		This relation is nevertheless kept unspecified as far as possible.
		Specifically, here we do not aim at a full derivation of thermodynamic quantities like Hawking temperature, horizon entropy, and specific heat -- which would require the study of quantum fields over a polymer black hole background or equivalently a saddle point analysis of the corresponding gravitational path integral -- but rather we are interested in studying the quantum corrections to classical thermodynamics coming from the effective quantum-corrected metric \eqref{metric}-\eqref{eq:asol}.
		Our strategy is then to start with the standard definitions of thermodynamic quantities for black holes \cite{CarlipBHThermodynamics,WaldBHthermodynamics,BardeenThefourlawsof,TownsendBlackHoles,HawkingTheLargeScale,WaldGRbook} and specify them for our static, quantum corrected metric.
		This can be heuristically motivated as follows.
		As we have discussed in Sec.~\ref{sec:effectiveQT}, quantum gravity effects are negligible in the low curvature regime so that already at the horizon the geometry is well approximated by the classical description plus sub-leading quantum corrections, at least for large masses for which the effective description is most reliable (cfr. Eqs.~\eqref{eq:brspapprox} and \eqref{eq:brsmapprox}).
		Therefore, one would expect also the thermodynamic properties of the quantum corrected horizon to be well approximated by the classical results at leading order.
		In particular, we do not assume a priori any entropy-area relation and instead compute the horizon entropy by integrating the first law of BH mechanics.
		Looking then at the large mass expansion of the resulting thermodynamical quantities, we can check to which extent the entropy-area law is satisfied.
		Moreover, being effective models of quantum black holes in general valid for macroscopic black holes, this will allow us also to compare and contrast our results with other effective models relying on different QG approaches. 
		\subsection{Temperature}\label{Sec:quantcorrectedtemperature}
\begin{figure}[t!]
			\centering
			\includegraphics[scale=0.7]{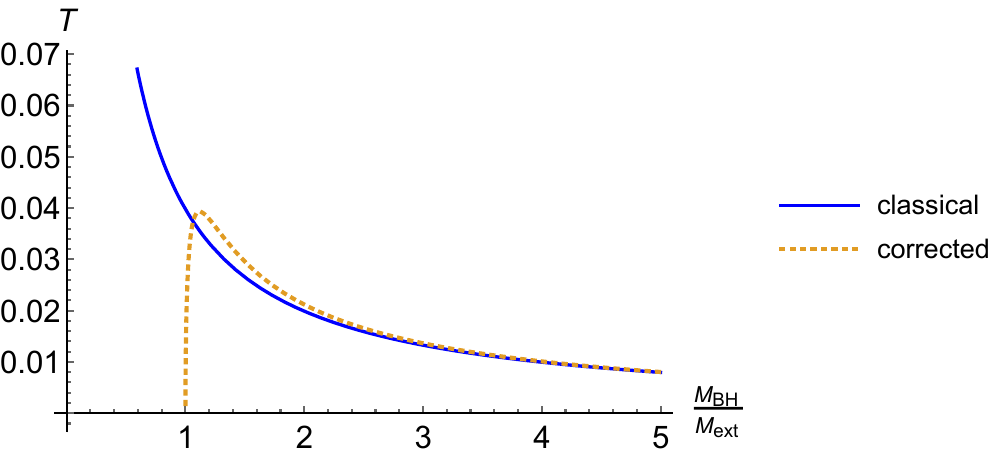}
			\caption{Plot of the temperature as a function of the BH mass for the quadratic mass relation Eq.~\eqref{eq:quadraticrelation} and $m = \lambda_k = 1$. The dotted line represents the quantum corrected result Eq.~\eqref{TBHfinal}, while the solid line is the classical result $T=1/8\pi M_{BH}$.
				The quantum result approaches the classical curve for large masses. At smaller masses the temperature reaches a maximum and becomes zero in the extremal configuration $M_{BH} = M_{ext}$. Similar behaviour is exhibited for other mass relations e.g. Eq. \eqref{symmassrelation}.}
			\label{Temperaturegraph}
		\end{figure}

\begin{figure}[t!]
			\centering
			\includegraphics[scale=0.575]{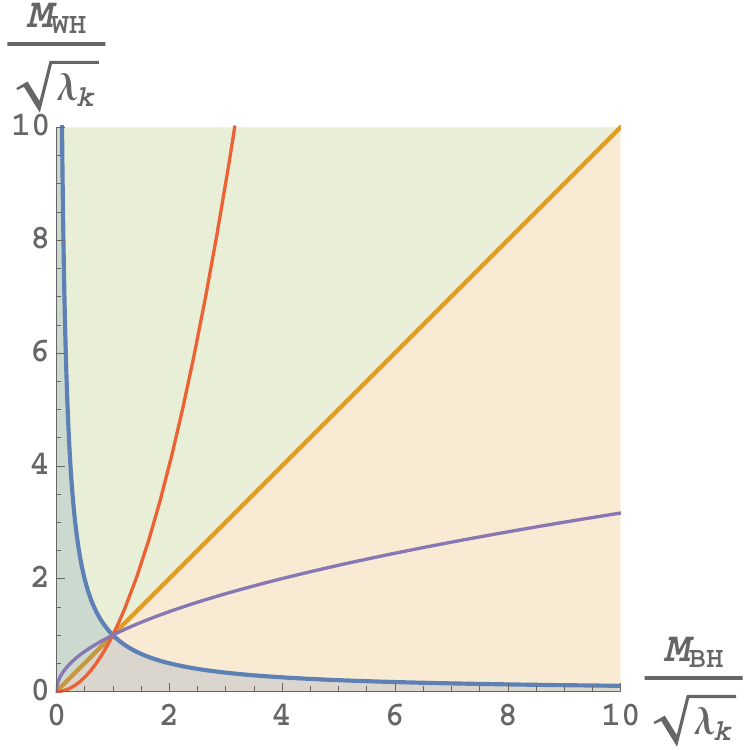}
			\caption{Plot of the extremal configuration $M_{BH} = \lambda_k/M_{WH}$ (blue hyperbola). Along this hyperbola in mass space both black and white hole horizons coincide, i.e. $x_+ = x_- = 0$ and $b(x_+) = b(x_-)$. Below this extremal configuration (blue shaded region) no horizons exist at all. The yellow line shows the symmetric mass relation $M_{BH} = M_{WH}$, along which $x_{\mathcal T} = 0$. In the green shaded region it is $x_{\mathcal T} > 0$, while in the yellow shaded region $x_{\mathcal{T}} < 0$. The red and violet lines correspond to the mass relation $M_{WH} = M_{BH}^2/\sqrt{\lambda_k}$ (red) and its inverse mass de-amplification $M_{WH} = M_{BH}^{1/2} \lambda_k^{1/4}$ (violet). They are chosen so that they intersect the extremal configuration when both masses are equal.}
			\label{fig:extmass}
		\end{figure}

		Assuming the no-hair theorems \cite{HawkingTheLargeScale,WaldGRbook,MisnerThorneWheeler} to be valid, effective quantum corrected black holes would be also characterized by only three macroscopic quantities:
		the mass, the angular momentum, and the charge.
		In our case, since the geometry is a quantum modification of the static, neutral, non-rotating Schwarzschild black hole, there is no charge nor angular momentum and the only relevant quantity is the mass.
		The First Law of black hole mechanics then reads 
		\be \label{eos}
		\dd M=T\dd S,
		\ee  
		where $M$ is the mass, $S$ is the horizon entropy, and $T$ is the Bekenstein-Hawking temperature.
		As anticipated above, we will compute the entropy by integrating Eq. \eqref{eos} as a function of the mass.
		The starting point for our thermodynamic analysis is then to compute the temperature as a function of the mass.
		To this aim, let us recall that the Bekenstein-Hawking temperature is related to the surface gravity via the relation
		\be \label{temperature}
		T=\frac{\kappa}{2\pi}\;,
		\ee 
		where the surface gravity $\kappa$ is defined by $\kappa^2=-g^{\mu\nu}g_{\rho\sigma}\nabla_{\mu}\chi^\rho\nabla_\nu\chi^\sigma/2=-g^{\mu\nu}g_{\rho\sigma}\Gamma^\rho_{\mu 0}\Gamma^\sigma_{\nu 0}/2$, with $\chi^\mu=(1,0,0,0)$ being a time-like Killing vector, and $\Gamma^{\mu}_{\nu\rho}$ being the Christoffel symbols associated to the connection compatible with the effective metric \eqref{metric}-\eqref{eq:asol}. The key feature that allows us to consider this is that we have a Killing horizon \cite{BardeenThefourlawsof,WaldGRbook,HawkingTheLargeScale}\footnote{Even though the original derivation makes use of a specific form of the Einstein field equations and one would expect that it holds only for general relativity, it is actually true for any static black hole and in addition for (1+1)-dimensional black holes \cite{WaldBlackHoleEntropy}.}.
		Following previous work on effective quantum corrected black holes derived from various approaches (see e.g. \cite{ModestoSelf-dualBlackHoles,NicoliniTdualBH}), to compute \eqref{temperature} it is convenient to bring our static spherically symmetric metric \eqref{metric} in the general form
		\be \label{bardeen}
		\dd s^2=-f(x)\dd v^2+2h(x)\dd v\dd x+b^2(x)\dd\Omega_2^2\;,
		\ee  
		\noindent
		for which the surface gravity at the horizon reads as
		\be\label{kappabardeen}
		\kappa^2=\left(\frac{f(x)'}{2h(x)}\right)^2\biggl|_{x=x_\pm}\;,
		\vspace{-0.15cm}
		\ee 
		\noindent
		where primes denote derivatives w.r.t. $r$, and $r_s$ denotes the radial coordinate of the horizon. As we have discussed in Sec. \ref{sec:effectiveQT}, our effective spacetime is characterized by two horizons respectively corresponding to the black and white hole sides. The computation of the thermodynamic quantities can be then specified for both horizons ($x=x_\pm$). In what follows we focus on the black hole side for which the horizon is located at $x=x_+$ given in Eqs.~\eqref{eq:xhorizon} and \eqref{eq:bhorizon}\footnote{By the symmetry arguments on the evolution through the Penrose diagram, we expect the corresponding analysis of the WH side to be the same with the role of black and white hole masses exchanged. In particular, for any given mass amplification or de-amplification, the WH thermodynamics should match the results of the BH side for the corresponding inverse mass de-amplification or amplification.}. 
		Recalling now from Sec.~\ref{sec:effectiveQT} (see below Eq. \eqref{eq:asympotic+}) that the coordinate rescaling necessary to recover the classical Schwarzschild solution in the asymptotic $x\to+\infty$ limit is given by  
		\be\label{asymptclassicalrescaling} 
		\tilde{r}=\left(\frac{8 \lambda_k M_{BH}^2}{M_{WH}}\right)^{\frac{1}{3}}\,x\qquad,\qquad\tau= \left(\frac{8 \lambda_k M_{BH}^2}{M_{WH}}\right)^{-\frac{1}{3}}\,t
		\vspace{-0.35cm}
		\ee 
		\noindent
		the resulting line element\footnote{Note that, as already stressed in the previous sections, $\lambda_j$ appears in the line element just as a coordinate rescaling. Hence, its precise value can not have any physical meaning and consistently in our later computations it will not appear in any physical quantity, which will in turn depend only on $\lambda_k$ and the mass observables.} 
		\begin{align}\label{temp1}
		\dd s^2=-\left(\frac{8 \lambda_k M_{BH}^2}{M_{WH}}\right)^{\frac{2}{3}}\frac{a(\tilde{r})}{\lambda_j^2}\dd\tau^2+\frac{\lambda_j^2}{a(\tilde{r})} \left(\frac{8 \lambda_k M_{BH}^2}{M_{WH}}\right)^{-\frac{2}{3}}\dd\tilde{r}^2+b^2(\tilde{r})\dd\Omega_2^2\;,
		\end{align}  
		\noindent
		can be brought into the form \eqref{bardeen} by using the null coordinate
		\be\label{eq:vdef}
		v=\tau+r^*\qquad\text{with}\qquad r^*=\int^{\tilde{r}}\frac{\dd\tilde{r}}{\mathfrak{a}(\tilde{r})}\quad,\quad \mathfrak{a}(\tilde{r}):=\left(\frac{8 \lambda_k M_{BH}^2}{M_{WH}}\right)^{\frac{2}{3}}\frac{a(\tilde{r})}{\lambda_j^2}
		\ee
		\noindent  
		such that $\dd v=\dd\tau+\dd\tilde r/\mathfrak{a}(\tilde{r})$ and we find
		\be
		f(\tilde r)=\left(\frac{8 \lambda_k M_{BH}^2}{M_{WH}}\right)^{\frac{2}{3}}\frac{a(\tilde{r})}{\lambda_j^2}\qquad,\qquad h(\tilde r)=1\;. 
		\vspace{-0.15cm}
		\ee
		\noindent
		The surface gravity at the horizon \eqref{kappabardeen} then becomes\footnote{The same result can be obtained by computing the periodicity of the Euclidean time coordinate resulting from Wick rotating the line element \eqref{temp1}. For completeness, the Euclidean analysis as well as the near-horizon geometry are reported in Appendix \ref{App:nearhorizon&Euclidean}.}
		\be\label{kappabardeenourbh}
		\kappa=\frac{1}{2}\left(\frac{8 \lambda_k M_{BH}^2}{M_{WH}}\right)^{\frac{2}{3}}\biggl|\frac{a'(\tilde r)}{\lambda_j^2}\biggr|\Biggl|_{\tilde{r}=\tilde{r}_s^{(+)}}\;,
		\ee
		\noindent
		where primes here denote derivatives w.r.t. $\tilde{r}$.
		Using now the chain rule and the expression \eqref{asymptclassicalrescaling} for $\tilde r$, we have
		\begin{equation}\label{temp2}
		\frac{1}{2}\left(\frac{8 \lambda_k M_{BH}^2}{M_{WH}}\right)^{\frac{2}{3}}\frac{a'(\tilde r)}{\lambda_j^2} = \frac{2}{\lambda_j^2}\left(\frac{\lambda_k M_{BH}^2}{M_{WH}}\right)^{\frac{1}{3}}\frac{\dd a}{\dd x}\;.
		\end{equation}
		\noindent
		Finally, using the horizon condition~\eqref{eq:xhorizon}, the derivative of $a(x)$ given in \eqref{eq:asol} evaluated at the BH horizon yields
		\be
		\frac{da}{dx}\bigg|_{x_+}=\lambda_j^2(1+x^2)\left(\frac{M_{BH}M_{WH}}{\lambda_k}\right)^{\frac{1}{3}}\frac{x}{(1+x^2)^{3/2}}\frac{1}{b^2(x)}\bigg|_{x_+}=\lambda_j^2\frac{x_+}{b^2(x_+)}\;,
		\ee
		\noindent
		so that, inserting the result together with \eqref{temp2} into \eqref{kappabardeenourbh}, we get 
		\be\label{TBHfinal}
		\kappa=\left(\frac{\lambda_k M_{BH}^2}{M_{WH}}\right)^{\frac{1}{3}}\frac{x_+}{b^2(x_+)}\qquad\overset{\eqref{temperature}}{\Longrightarrow}\qquad T=\frac{1}{2\pi}\left(\frac{\lambda_k M_{BH}^2}{M_{WH}}\right)^{\frac{1}{3}}\frac{x_+}{b^2(x_+)}\;.
		\ee
		\noindent
		The temperature \eqref{TBHfinal} is shown in Fig.~\ref{Temperaturegraph} for the mass relation Eq.~\eqref{eq:quadraticrelation}, which at this stage has been chosen just as a representative for plotting purposes. Similar behavior is obtained for other mass relations. 
		Independently of the specific mass relation, for large masses $M_{BH}M_{WH}\gg\lambda_k$, $x_+\simeq(M_{BH}M_{WH}/\lambda_k)^{1/3}$ (cfr. Eq.~\eqref{eq:xhorizon}) and $b(x_+)\simeq 2M_{BH}$ (cfr. Eq. \eqref{eq:brspapprox}) so that the classical result $T\simeq\frac{1}{8\pi M_{BH}}$ is recovered.
		On the other hand, unlike the classical result, which diverges for the horizon radius approaching the Planck regime, the quantum corrected temperature reaches a maximum value where quantum effects get relevant and, continuing to smaller sizes, cools down and vanishes when $M_{BH}$ reaches the extremal configuration $M_{BH} M_{WH}(M_{BH})=\lambda_k$ ($x_+=0$ cfr. Eq. \eqref{eq:xhorizon}).
		In this scenario both horizons coincide and an extremal black hole is reached.
		We define the extremal mass value implicitly via
		\begin{equation}\label{implicitdefextremalmass}
		M_{ext} \, M_{WH}(M_{ext}) = \lambda_k \;.
		\end{equation}
		\noindent
		Of course $M_{ext}$ depends then on the choice of a mass relation $M_{WH}(M_{BH})$.
		This is clear as the extremal situation is a one-dimensional hyperbola in ``mass space'', spanned by $M_{BH}$ and $M_{WH}$, see Fig.~\ref{fig:extmass}.
		$M_{ext}$ is then the black hole mass for which $M_{WH}$ intersects the extremal hyperbola for which the horizon reaches its minimal possible size for this mass relation.
		An interesting question is then whether the horizons and the transition surface coincide in the extremal situation.
		It is easy to check that only in the symmetric case $M_{BH} = M_{WH}$, $x_{\mathcal T } = x_+ = x_- = 0$.
		For all other scenarios this is not true.
		The situation is depicted in Fig.~\ref{fig:extmass}.
		Nevertheless, this is only a statement at the extremal configuration and therefore fixes the value of the mass relating function only in one point.
		Transition surface and the extremal horizon coincide whenever both masses are equal at the extremal configuration, i.e. $M_{BH} = M_{WH} = M_{ext} = \sqrt{\lambda_k}$.
		This means for $M_{WH} = \mathcal F(M_{BH})$, $\mathcal F$ needs to satisfy the condition $\mathcal F(M_{ext}) = \sqrt{\lambda_k}$.
		In the case of the quadratic mass relation Eq.~\eqref{eq:quadraticrelation} this condition fixes the value of $m$ in terms of $\lambda_k$, namely $m=\sqrt{\lambda_k}$. For such a value of $m$, the critical length scale $\ell_{crit} = \left(8 \lambda_k m\right)^\frac{1}{3}$ (cfr. Eq. \eqref{eq:classregimeb+quadratic}) becomes $\ell_{crit}=2\sqrt{\lambda_k}$ which is nothing but twice the minimal areal radius of the extremal black hole configuration as can be easily seen from the general expression \eqref{eq:bhorizon} yielding $b^2(r_s^{(\pm)})=\lambda_k=:b_{ext}^2$ when $M_{BH} = M_{WH} = M_{ext} = \sqrt{\lambda_k}$. This is compatible with the expectation of the onset of quantum effects occurring deep in the high curvature/small length regime close to the resolved singularity as discussed in Sec. \ref{sec:effectiveQT}.
		Whether or not the parameter $m$ needs to be fixed in such a way remains open and further investigation would be needed. In this respect, we remark that the present analysis is based on effective models, where it is (implicitly) assumed that there exists a quantum state which is always peaked on the effective geometry.
		Up to now, there is no evidence of whether this is the case or not.
		In fact, it is rather likely, that this is not the case for small masses.
		Therefore the analysis of small masses remains heuristic at this stage and a proper understanding of the deep quantum regime requires a full development of a polymer Hilbert space quantum theory and the corresponding semi-classical sector which is left for future work.
		
		Nevertheless, remaining at the effective level, let us notice that as discussed also in other models \cite{ModestoSelf-dualBlackHoles,NicoliniTdualBH,NicoliniHolographicScreensin} such an extremal configuration corresponds to an extremal non-radiating small object of genuine quantum gravitational character, which in our case can be thought of as a remnant of mass $M_{BH}=\frac{\lambda_k}{M_{WH}}$ in a similar spirit to previous LQG proposals \cite{BianchiWhiteHolesas}.
		Finally, to make the comparison with the classical case and the role of quantum corrections more transparent, let us consider the large mass expansion of the temperature.
		To this aim, recalling the full expression \eqref{eq:bhorizon} for the horizon areal radius, Eq. \eqref{TBHfinal} can be written more explicitly as
		\be\label{temp}
		T=\frac{M_{BH}}{\pi}\frac{\sqrt{1-y^2}}{\left(M_{BH}^2f(y)^3+\frac{\lambda_k^2}{M_{BH}^2}\frac{1}{f(y)^3}\right)}\;,
		\vspace{-0.15cm}
		\ee  
		\noindent
		where we introduced the shorthand notations
		\be\label{criticalmass}
		y:=\left(\frac{\lambda_k}{M_{BH}M_{WH}}\right)^{\frac{1}{3}}\qquad,\qquad
		f(y):=1+\sqrt{1-y^2}\;.
		\ee
		\noindent
		Expanding then \eqref{temp} around $y=0$, which corresponds to a large $M_{BH}M_{WH}$ expansion (i.e. $M_{BH}M_{WH}\gg\lambda_k$), we get  
		\be\label{Tlargemassexpansion}
		T=\frac{1}{8\pi M_{BH}}\left[1+\frac{y^2}{4}+\mathcal O\left(\frac{M_{WH}^2}{M_{BH}^2}y^3\right)\right]\;,
		\ee
		from which we see that the classical result is reproduced at leading order and quantum corrections go to zero for large masses or equivalently as $\lambda_k\to0$, for all mass relations\footnote{This means for all mass relation for which $M_{BH} M_{WH} \gg \lambda_k$ is satisfied for large black hole masses. This excludes for instance the case $M_{WH}\propto 1/M_{BH}$ or generic inverse power relations for which sufficiently large black holes would correspond to microscopic white holes thus casting doubts on the validity of the effective description on the WH side.}.
		Moreover, consistently with the observation that the horizon size is slightly smaller due to quantum corrections, the first-order correction here is positive, i.e. the black hole is slightly hotter than expected classically as reflected also by the plots in Fig. \ref{Temperaturegraph} with the quantum-corrected result approaching the classical one from above as the mass increases.
		\subsection{Specific Heat}\label{sec:specificheat}
		To compute the specific heat $C=\frac{\dd M_{BH}}{\dd T}$, we can differentiate the expression for the quantum corrected temperature w.r.t. to $M_{BH}$ and then reverse the result. To this aim, we need to take into account the fact that the mass dependence occurs in the expression \eqref{temp} both explicitly via $M_{BH}$ and implicitly into $y$ via $M_{WH}$ (cfr. Eq. \eqref{criticalmass}), the latter being thought of as a generic function $M_{WH}(M_{BH})$. Keeping then the relation between $M_{WH}$ and $M_{BH}$ unspecified, we have
		\be\label{implicit1overC}
		\frac{1}{C}=\frac{\dd}{\dd M_{BH}}T\Bigl(M_{BH},y(M_{BH})\Bigr)=\frac{\partial T}{\partial M_{BH}}+\frac{\partial T}{\partial y}\frac{\partial y}{\partial M_{BH}}\;,
		\ee
		\noindent
		with
		\be
		\frac{\partial y}{\partial M_{BH}}=\frac{\partial}{\partial M_{BH}}\left(\frac{\lambda_k}{M_{BH}M_{WH}}\right)^{\frac{1}{3}}=-\frac{1}{3}\frac{\lambda_k^{1/3}}{(M_{BH}M_{WH})^{4/3}}\left(M_{WH}+M_{BH}\frac{\partial M_{WH}}{\partial M_{BH}}\right)\;.
		\ee
		\noindent
		Thus, using the expression \eqref{temp} of $T$ to explicitly compute the derivatives in \eqref{implicit1overC}, we get
		\be\label{1overCfinal}
		\frac{1}{C}=\frac{T}{M_{BH}}\left[1-2\frac{G^-(y)}{G^+(y)}+\frac{1-\sqrt{1-y^2}}{1-y^2}\left(\frac{f(y)}{3}-\sqrt{1-y^2}\frac{G^-(y)}{G^+(y)}\right)\left(1+\frac{M_{BH}}{M_{WH}}\frac{\p M_{WH}}{\p M_{BH}}\right)\right]
		\ee
		\noindent
		where 
		\be
		G^{\pm}(y):=M_{BH}^2f(y)^3\pm\frac{\lambda_k}{M_{BH}^2f(y)^3}\;.
		\ee
		\begin{figure}[t!]
			\centering 
			\includegraphics[scale=0.7]{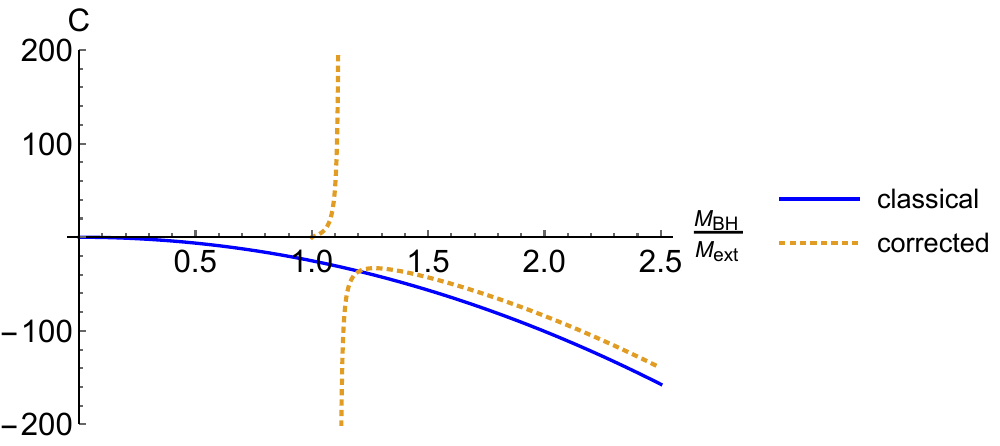}
			\caption{Plot of the specific heat as a function of the BH mass for the mass relation Eq.~\eqref{eq:quadraticrelation}. The dotted line represents the quantum corrected result, while the solid line is the classical result $C=-8\pi M_{BH}^2$. As before, the quantum corrected plot refers to the mass relation $M_{WH}=M_{BH}^2/m$ and we set $\lambda_k=m=1$. The discontinuity where the quantum corrected specific heat changes sign is interpreted as a phase transition of the system.}
			\label{Ccomparison}
		\end{figure}
		The behaviour of $C$ as a function of $M_{BH}$ for the mass relation Eq.~\eqref{eq:quadraticrelation} is shown in Fig.~\ref{Ccomparison}. As we can see from the plot, for small masses, the quantum corrected curve significantly deviates from the classical one.
		Indeed, a divergence occurs when the temperature reaches its maximum ($\dd T/\dd M_{BH}=0$ and hence $1/C$ in \eqref{implicit1overC} also vanishes).
		Such a discontinuity in the specific heat separates a regime in which $C<0$ from a regime in which $C\geq0$ with $C=0$ for the extremal configuration $M_{BH}=\lambda_k/M_{WH}$ and can be interpreted as signaling a phase transition from large thermodynamically unstable black holes to small stable quantum black holes \cite{CarlipBHThermodynamics,ModestoSelf-dualBlackHoles,NicoliniTdualBH,NicoliniHolographicScreensin}.
		Although this is not explicitly visible from Fig.~\ref{Ccomparison}, where we focused on the quantum corrected small mass region, for large masses, the classical behavior is approached.
		This can also be seen by expanding \eqref{1overCfinal} around $y=0$ (or equivalently $M_{BH}M_{WH}\gg\lambda_k$) thus yielding
		\be\label{eq:Cexp}
		C\simeq-8\pi M_{BH}^2\left[1-\frac{1}{6}\left(\frac{5}{2}+\frac{M_{BH}}{M_{WH}}\frac{\p M_{WH}}{\p M_{BH}}\right)y^2+\dots\right]\;,
		\ee
		\noindent
		which is again the classical result plus quantum corrections going to zero as $\lambda_k\to0$.
		Nevertheless, the first order correction is directly dependent on the derivative of the mass relation.
		It has to be assumed that this coefficient remains subdominant, which is the case for reasonable mass relations (say $M_{WH} \propto (M_{BH})^\varepsilon$, $\varepsilon \geq 0$).
		\subsection{Entropy: Mass Relation, Minimal Length, and Logarithmic Corrections}\label{sec:entropy}
\begin{figure}[t!]
			\centering
			\subfigure[]
			{\includegraphics[scale=0.51]{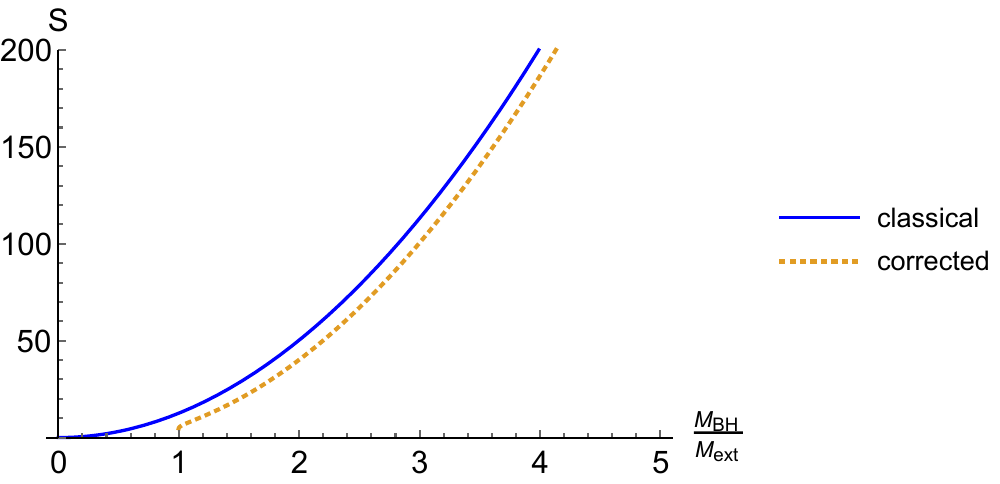}}
			\subfigure[]
			{\includegraphics[scale=0.525]{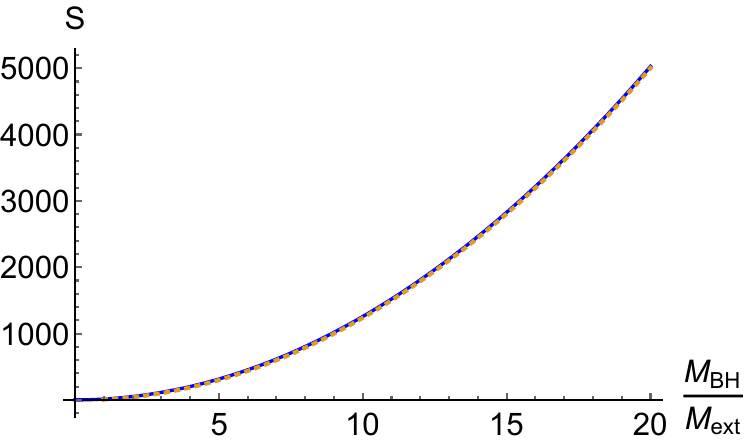}}
			\caption{Plot of the horizon entropy as a function of the BH mass. The quantum corrected entropy (dashed orange line) vanishes for $M_{BH}=M_c$ (a), while it approaches the classical result (solid blue line) for large masses (b). For the quantum corrected plot we choose $M_{WH}=M_{BH}^2/m$ and the parameters $\lambda_k=m=1$.}
			\label{BHentropyplot}
		\end{figure}

		We can finally proceed to compute the entropy of the quantum corrected black hole by integrating the first law \eqref{eos}.
		This provides us with the expression of the BH entropy in terms of the ADM energy which in the case under consideration amounts to
		\be\label{entropyfrom1stlaw}
		S=\int_{M_{ext}}^{M_{BH}}\frac{\dd M_{BH}}{T}\;.
		\ee
		\noindent
		The evaluation of the integral \eqref{entropyfrom1stlaw} with $T(M_{BH})$ given in \eqref{TBHfinal} or equivalently \eqref{temp} requires a fixed mass relation $M_{WH}\left(M_{BH}\right)$ in order to make the integral well-posed.
		Furthermore, the expression of the temperature \eqref{TBHfinal} (equiv. \eqref{temp}) is too complicated for analytical computations, and therefore the integral is solved numerically.
		The result for the mass relation Eq.~\eqref{eq:quadraticrelation} is reported in Fig. \ref{BHentropyplot}.
		As we can see from the plot, the classical behaviour is approached for large masses (cfr. Fig. \ref{BHentropyplot} (b)), while by construction the entropy of the quantum corrected black hole becomes zero for $M_{BH}=M_{ext}$ (cfr. Fig. \ref{BHentropyplot} (a)).
		Combining this with the results of Sec. \ref{Sec:quantcorrectedtemperature}, we conclude that the extremal configuration is then characterised by vanishing temperature and vanishing entropy, thus confirming its interpretation as a thermodynamically stable extremal remnant of the black hole life cycle shared also by other approaches to quantum BHs \cite{ModestoSelf-dualBlackHoles,NicoliniTdualBH,NicoliniHolographicScreensin}.
		Despite a full analytical solution not being available for the integral \eqref{entropyfrom1stlaw}, in the limit of large masses $M_{BH} M_{WH} \gg \lambda_k$, the expression for the temperature can be integrated order by order.
		Specifically, recalling the expansion \eqref{Tlargemassexpansion}, the integral \eqref{entropyfrom1stlaw} can be written as
		\begin{align}
		S&\simeq\int_{M_{ext}}^{M_{BH}}\dd M_{BH}\,8\pi M_{BH}\left(1-\frac{y^2}{4}+\frac{y^6}{64}+\dots\right)\nonumber\\
		&=4\pi(M_{BH}^2-M_{ext}^2)-2\pi\int_{M_{ext}}^{M_{BH}}\dd M_{BH}\left(\frac{\lambda_k^2M_{BH}}{M_{WH}^2}\right)^{\frac{1}{3}}+\frac{\pi}{8}\int_{M_c}^{M_{BH}}\dd M_{BH}\left(\frac{\lambda_k^2}{M_{BH}M_{WH}^2}\right)+\cdots\label{entropymassexp}
		\end{align}
		\noindent
		from which we see that as expected the classical result is obtained at leading order, with $4\pi r_s^2=16\pi M_{BH}^2$ (resp. $4\pi r_{ext}^2=16\pi M_{ext}^2$) being the classical horizon area of a black hole of mass $M_{BH}$ (resp. $M_{ext}$ which goes to zero in the classical limit $\lambda_k\to0$).
		Higher-order terms yield quantum corrections that go to zero in the classical limit and the relative errors w.r.t. the leading order gets smaller as the mass increases.
		It is interesting to note that the explicit form of the next-to-leading-order corrections to the entropy \eqref{entropymassexp} are completely determined once a relationship between the black and white hole masses is specified.
		Therefore, depending on the imposed mass relation, it is possible to obtain different correction laws.
		In particular, it would be possible to match the results of different quantum models of black holes by simply modifying the mass relation.
		In turn, this affects the onset of quantum effects as there are always quantum effects driven by the curvature, but a mass-relation dependent critical length scale (cfr. Eqs.~\eqref{eq:classregimeb+} and \eqref{eq:classregimeb-}).
		Unfortunately, from what concerns purely the study of effective models, it is not clear how a certain mass relation gives a microscopic explanation for a certain entropy correction.
		This is because the present model is effective and no microscopic structure is available.
		The situation might change once a full quantum theory is constructed.
		The model then still remains restrictive as there is a strong symmetry assumption imposed already at the classical level.
		Nevertheless, in line with the aim of the present investigation, at this stage with the quantum corrected entropy \eqref{entropymassexp} at our disposal it is possible to analyze the diagnostic of different mass relations and ask what kind of phenomenological consequences they might correspond to. For instance, a symmetric mass relation $M_{BH} = M_{WH}$ yields a first-order correction of the form $\propto M_{BH}^{2/3}$.
		A correction $\propto M_{BH}$ as deduced within the framework of non-commutative geometry \cite{AnacletoNoncummutativecorrectionto} can instead be reproduced by the mass relation $M_{BH} = M_{WH}^2/m$, $[m] = M$, i.e. $M_{WH}=(m M_{BH})^{1/2}$.
		However, this is only true for the first-order correction and does not reproduce higher corrections.
		More interestingly, among the various possible mass relations, we notice that for a quadratic mass relation (see Eq.~\eqref{eq:quadraticrelation}) of the kind $M_{WH}=M_{BH}^2/m$ with $[m]=M$, Eq.~\eqref{entropymassexp} gives
		\be\label{Entropylogcorrection1}
		S=4\pi(M_{BH}^2-M_{ext}^2)-2\pi\left(m\lambda_k\right)^{\frac{2}{3}}\log\left(\frac{M_{BH}}{M_{ext}}\right)-\mathcal O\left(\frac{m^2\lambda_k^2}{M_{BH}^4}\right)\;,
		\ee
		\noindent
		or written in terms of the horizon area and the critical length $\ell_{crit}=(8m\lambda_k)^{1/3}$
		\be\label{Entropylogcorrection2}
		S\simeq\frac{1}{4}(A-A_{ext})-\frac{\pi}{4}\ell_{crit}^2\log\left(\frac{A}{A_{ext}}\right)\;.
		\ee
		
		\noindent
		Therefore, for such a quadratic mass relation, the quantum corrections are in agreement with the expectations of a logarithmic term as a next to leading order correction to the classical Bekenstein-Hawking area law from all the major approaches to quantum gravity, both at the effective and full theory level \cite{StromingerMicroscopicOriginOf,MannUniversalityofQuantum,RovelliBHentropyfromLQG,AshtekarQuantumGeometryAndBlackHoleEntropy, AshtekarQuantumgeometryof,KaulQuantumBHEntropy,KaulLogarithmicCorrectionto,CarlipLogarithmicCorretionsTo,BarberoQuantumisolatedhorizons,MeissnerBlackholeentropy,LivineQuantumblackholes,SenBHEntropyFunction,CaravelliHolographiceffectiveactions,JeonLogarithmiccorrectionsto}. 
		It is quite remarkable that for our model this is the case exactly for the special kind of mass relation discussed at the end of Sec.~\ref{sec:effectiveQT} (cfr. Eq.~\eqref{eq:quadraticrelation}) for which there is a strict distinction between high curvature and small length quantum effects originating from the $k$- and the $j$-polymerisation, respectively.
		This seems to indicate a direct link between finite-length quantum effects and logarithmic entropy corrections\footnote{We have cross-checked if this same relation between finite-length effects and logarithmic corrections also appear in similar models, to be precise the model \cite{BodendorferEffectiveQuantumExtended}. As there are also two independent mass observables, it is possible to find a mass relation that gives logarithmic corrections to the entropy. Nevertheless, for the model \cite{BodendorferEffectiveQuantumExtended}, this mass relation cannot be interpreted as related to pure finite-length effects. In particular, it is not in the class of mass relations leading to a universal curvature bound within this model (see \cite{BodendorferEffectiveQuantumExtended} for details).}. In this respect, let us notice that for the model under consideration here the occurrence of proper finite-length effects for the quadratic mass relation is compatible with a minimal length generalized uncertainty principle (GUP) associated with the existence of a critical length scale for the onset of quantum effects.
		To see this, let us first go back to the expression of the temperature and specify it for the quadratic mass relation \eqref{eq:quadraticrelation}. In this case, taking into account the implicit definition \eqref{implicitdefextremalmass} of the extremal mass and the expression \eqref{criticalmass} of $y$, we have 
		\be\label{quadraticmass:critical}
		M_{ext}=(m\lambda_k)^{\frac{1}{3}}=\frac{\ell_{crit}}{2}\qquad,\qquad y=\frac{M_{ext}}{M_{BH}}
		\ee
		\noindent
		and the expression \eqref{Tlargemassexpansion} for the temperature yields
		\be\label{TBHquadraticmassrelation}
		T=\frac{1}{8\pi M_{BH}}\left[1+\frac{M_{ext}^2}{4M_{BH}^2}+\mathcal O\left(\frac{M_{ext}^4}{M_{BH}^4}\right)\right]\;.
		\ee
		\noindent
		Consider now a GUP of the form\footnote{This form of GUP has been suggested for instance in string theory \cite{VenezianoAStringyNature,GrossStringTheoryBeyond}, polymer quantization motivated by LQG \cite{HossainBackgroundIndependentGUP}, or from general consideration about quantum mechanics and gravity as e.g. in \cite{MaggioreGUP,AdlerOnGravityAnd,ScardigliGUP}, and references within. We should mention however that modifications of the GUP in the Planckian or sub-Planckian regime have also been proposed in the literature (see e.g. \cite{CarrGUPselfdualBH,CarrBHGUP} where in turn the modifications were motivated by the loop quantum black hole of \cite{ModestoLoopquantumblack,ModestoBlackHoleinterior,ModestoSelf-dualBlackHoles,ModestoSemiclassicalLoopQuantum}). Here, for concreteness, we focus on \eqref{GUPexample} as we are only interested in comparing the leading corrections in the large mass expansion of thermodynamic quantities (specifically the temperature which yields the logarithmic term in the entropy). The small mass behaviors of the corresponding temperatures are different thus suggesting that similar modifications of the GUP might occur also in our model below $\ell_{crit}$.}
		\be\label{GUPexample}
		\Delta q\Delta p\geq 1+\alpha^2(\Delta p)^2\;,
		\ee
		\noindent
		where we set $\hbar=1$, we denoted the position uncertainty by $\Delta q$ rather than $\Delta x$ to avoid confusion with the $x$-notation used in previous sections for the radial coordinate, and $\alpha$ is a dimensional parameter of the order of the Planck scale that will be specified below. The above uncertainty principle can be used to compute a quantum corrected Hawking temperature associated with the characteristic energy $\Delta p$ ($c=1$) of the emitted Hawking-pair photons. Specifically, from \eqref{GUPexample}, we have
		\be
		\Delta p=\frac{\Delta q}{2\alpha^2}\left(1-\sqrt{1-\frac{4\alpha^2}{(\Delta q)^2}}\,\right)\;,
		\ee
		and, for $\Delta q=2M_{BH}(\simeq b(r_s^{(+)})$ for sufficiently large BH masses), a $M_{BH}\gg\alpha$ expansion yields \cite{ChenBHRemnantsDarkMatter,ChenGUPDarkMatter}
		\be\label{TGUPlargemass}
		T_{\text{GUP}}=\epsilon\,\Delta p=\frac{\epsilon}{2M_{BH}}\left[1+\frac{\alpha^2}{4M_{BH}^2}+\mathcal O\left(\frac{\alpha^4}{M_{BH}^4}\right)\right]\;,
		\ee
		\noindent
		where $\epsilon$ denotes a calibration parameter. The expression \eqref{TGUPlargemass} reduces to \eqref{TBHquadraticmassrelation} by setting $\epsilon=1/4\pi$ and identifying $\alpha=M_{ext}$.
		Moreover, given a GUP of the form \eqref{GUPexample}, the allowed minimal position uncertainty $\Delta q_{min}$ is determined by the conditions
		\be
		\begin{cases}
			F(\Delta q,\Delta p)=0\\
			\frac{\partial}{\partial\Delta p}F(\Delta q,\Delta p)=0
		\end{cases}\qquad\text{with}\qquad F(\Delta q,\Delta p):=\Delta q\Delta p-1-\alpha^2(\Delta p)^2\;,
		\ee
		\noindent
		which yield the solution
		\be
		\Delta q_{min}=2\alpha=2M_{ext}=\ell_{crit}\;,
		\ee
		\noindent
		compatible with the results of Sec. \ref{sec:effectiveQT} according to which, in the case of a quadratic relation between $M_{WH}$ and $M_{BH}$, the $j$-sector polymerisation yields a critical length scale below which small length quantum effects become dominant. In particular, as discussed in Sec. \ref{Sec:quantcorrectedtemperature} (cfr. Fig. \ref{fig:extmass} and discussion below Eq. \eqref{implicitdefextremalmass}), if $m=\sqrt{\lambda_k}=M_{ext}$ as determined by the horizon cross section of the extremal minimal-sized configuration to coincide with that of the transition surface, then $\Delta q_{min}=\ell_{crit}=2\sqrt{\lambda_k}=2b_{ext}$. Finally, a comparison of the temperature \eqref{TBHquadraticmassrelation} or \eqref{TGUPlargemass} with the GUP literature (see e.g. \cite{ChenBHRemnantsDarkMatter,ChenGUPDarkMatter}), where a GUP of the form \eqref{GUPexample} yields a leading correction of order $M_{Pl}^2/M_{BH}^2$ with $M_{Pl}$ denoting the Planck mass, would in principle allow us to relate the extremal mass $M_{ext}$ (hence $\lambda_k$) to the Planck mass as $\alpha=M_{ext}=2M_{Pl}$ in agreement with our expectation of the minimal extremal BH configuration being Planck-sized.
		
		In a similar spirit, further comparison with the existing literature sharing a logarithmic first-order correction to the entropy can be useful to constraint some free parameters in the present model. In particular, several results based on different approaches predict the numerical coefficient in front of the logarithmic correction to be $-3/2$. This is for instance the case of \cite{CarlipLogarithmicCorretionsTo} where a quite general argument based on CFT techniques applied to asymptotic or near-horizon symmetries was used to compute the first-order quantum correction to the Cardy entropy formula. The same coefficient is obtained also in the context of loop quantum gravity from state-counting arguments based on combinatorics of boundary states of a $SU(2)$ Chern-Simons theory on the horizon \cite{Agullo:2009eq,Engle:2011vf,KaulLogarithmicCorrectionto}. In both cases, the leading quantum correction to the classical entropy-area relation is found to be
		\be 
		S\simeq\frac{A}{4 l_p^2}-\frac{3}{2}\log\left(\frac{A}{l_p^2}\right),
		\ee 
		where $l_p$ is the Planck length. Comparing it with our result of the large mass expansion of the entropy, we see that the logarithm coefficients coincide for the specific value of critical length $\ell_{crit}=\sqrt{\frac{6}{\pi}}$ which, for $m=\sqrt{\lambda_k}=M_{ext}$ as determined by the horizon cross-section of the extremal minimal-sized configuration to coincide with that of the transition surface, amounts in turn to $ \lambda_k=\frac{3}{2\pi}$ in Planck units. Of course, such comparison is only qualitative and further work would be needed to eventually relate our effective treatment with a specific microscopic quantum model.
		
		\subsection{Evaporation Time}\label{sec:evaporationtime}
		
		With all previous thermodynamic quantities of the quantum black hole at our disposal, it is also possible now to compute its lifetime according to Hawking radiation.
		Let us assume then a Stefan-Boltzmann law for black body radiation (see \cite{WaldGRbook} for semi-classical black hole evaporation or \cite{ModestoLoopquantumblack,ModestoEvaporatingLBH,ModestoBlackHoleinterior,ModestoSelf-dualBlackHoles,ModestoSemiclassicalLoopQuantum}
		for LQG black holes)
		\begin{equation}
		\frac{\dd M_{BH}}{\dd v} = \mathcal L = -\sigma A(M_{BH}) \, T^4(M_{BH}) \;,
		\end{equation}
		
		\noindent
		where $v$ is the ingoing Eddington-Finkelstein time (see Eq.~\eqref{eq:vdef}), $\sigma$ the Stefan-Boltzmann constant, $A=4\pi b^2(x_+)$ the horizon area with areal radius $b$ given in \eqref{eq:bhorizon}, and the temperature $T$ is given in Eq. \eqref{TBHfinal} or \eqref{temp}.
		Note that $v$ coincides with the Schwarzschild time $t$ at $r \rightarrow \infty$.
		To integrate the above expression it is convenient to rewrite the luminosity $\mathcal L$ as
		\begin{align}
		\nn-\mathcal L =& \frac{\sigma}{4 \pi^3} \frac{M_{BH}^4}{P^4} \frac{x_+^4}{b^6(x_+)} \\
		=& \frac{\sigma}{4 \pi^3} \frac{M_{BH}^4}{P^4} \frac{\left(P^2-1\right)^2}{\lambda_k^3 \cosh^3\left(\ln\left(Q\right)+3 \ln\left(P+\sqrt{P^2-1}\right)\right)} \,,
		\end{align}
		
		\noindent
		where $P = 1/y = (M_{BH} M_{WH}/\lambda_k)^{1/3}$, $Q = M_{BH}/M_{WH}$, and we used the relation (cfr. Eq. \eqref{eq:bhorizon})
		\begin{align}
		b^2(x_+)&=\frac{\lambda_k}{2}\left(Q(P+\sqrt{P^2-1})^3+\frac{1}{Q(P+\sqrt{P^2-1})^3}\right)\nonumber\\
		&=\frac{\lambda_k}{2}\left(e^{\ln(Q(P+\sqrt{P^2-1})^3)}+e^{-\ln(Q(P+\sqrt{P^2-1})^3)}\right)\nonumber\\
		&=\lambda_k\cosh\left(\ln(Q)+3\ln\left(P+\sqrt{P^2-1}\right)\right)\;.
		\end{align}
		
		\noindent
		The evaporation time is then given as the mass integral
		\begin{equation}
		v= \frac{4 \pi^3 \lambda_k^3}{\sigma} \int_{M_{init}}^{M_{final}}\dd M_{BH} \frac{P^4}{M_{BH}^4} \frac{\cosh^3\left(\ln\left(Q\right)+3 \ln\left(P+\sqrt{P^2-1}\right)\right)}{\left(P^2-1\right)^2} \,,
		\end{equation}
		
		\noindent
		where again a mass relation $M_{WH} = M_{WH}\left(M_{BH}\right)$ has to be assumed and of course then $M_{WH}$, $Q$, and $P$ are functions of $M_{BH}$ only.
		For the above quadratic mass relation \eqref{eq:quadraticrelation}, i.e. $M_{WH} = M_{BH}^2/m$, and further $m = \sqrt{\lambda_k} = M_{ext}$, we find 
		\begin{align}
		Q =& \frac{M_{BH}}{M_{WH}} = \frac{\sqrt{\lambda_k}}{M_{BH}} =: \frac{1}{z} \,,
		\\
		P =& \left(\frac{M_{BH} M_{WH}}{\lambda_k}\right)^\frac{1}{3} = \frac{M_{BH}}{\sqrt{\lambda_k}} =: z \,.
		\end{align}
		
		\noindent
		The evaporation time is then
		\begin{equation}
		v(z_i,z_f)= \frac{4 \pi^3 \sqrt{\lambda_k}^3}{\sigma} \int_{z_{i}}^{z_{f}}\dd z \frac{\cosh^3\left(-\ln\left(z\right)+3 \ln\left(z+\sqrt{z^2-1}\right)\right)}{\left(z^2-1\right)^2} \,.
		\end{equation}
		
\begin{figure}[t!]
			\centering 
			\includegraphics[scale=0.7]{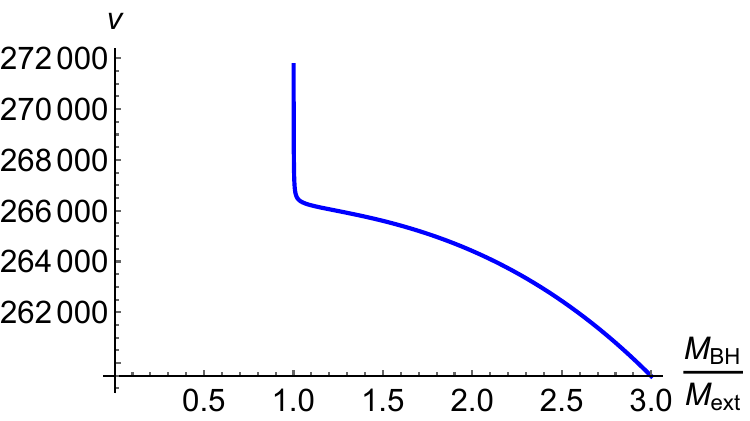}
			\caption{Plot of the evaporation time $v(z_i,z_f)$ for a large initial black hole $z_i = 10 \gg 1$. The evaporation time $v$ diverges for $M_{BH} \rightarrow M_{ext}$. Here $M_{ext} = \sqrt{\lambda_k}$, where the constants are chosen as $\sigma = 1$, $\lambda_k = 1$.}
			\label{evaporation}
		\end{figure}

		This expression can be analytically integrated with help of \texttt{Mathematica} or similar computer algebra software.
		The final expression is complicated and not very insightful.
		Fig.~\ref{evaporation} depicts the evaporation time as a function of the final state $z_f$.
		It is clearly visible that $v(z_i,z_f) \xrightarrow{z_f \rightarrow 1} \infty$, i.e. it takes infinite $v$-time to reach the extremal configuration $M_{BH}=M_{ext}$ ($z_f = 1$).
		For macroscopic black holes, i.e. $z_i \gg 1$, and $z_f \sim 1$, the evaporation time can be approximately written as 
		\begin{equation}
		v(z_i,z_f) \simeq \frac{\sqrt{\lambda_k}^3 \pi}{\sigma} \frac{1}{z_f-1} - \frac{26 \sqrt{\lambda_k}^3 \pi}{\sigma} \ln\left(z_f-1\right) + \frac{256 \sqrt{\lambda_k}^3 \pi}{3 \sigma} z_i^3 + \mathcal{O}\left(\sqrt{z_f-1}\right) + \mathcal{O}\left(z_i\right) \;.
		\end{equation}
		
		\noindent
		Note that the last term is exactly the classical contribution, i.e. 
		\be
		v_{evap}^{cl} = \frac{256 \sqrt{\lambda_k}^3 \pi}{3 \sigma} z_i^3 = \frac{256\pi}{3 \sigma} M_i^3 \;.
		\ee
		
		\noindent
		The divergence of the evaporation time can thus be quantified as of order $1/(z_f-1)$ and a logarithmic divergence $-\ln(z_f-1)$, while the classical result is obtained for large initial black hole masses and neglecting the divergent terms.
		
		The divergence of the evaporation time is consistent with the fact that the specific heat converges to zero (cfr. Eq.~\eqref{1overCfinal} and Fig.~\ref{Ccomparison}).
		The ``speed'' of evaporation thus slows down with decreasing mass.
		This shows that the black hole reaches asymptotically an extremal remnant state, which is thermodynamically stable.
		
		Few comments are however in order. Let us notice that the above computation is based on at least two approximations both of which can be questioned or in principle even broken.
		The first one is that the computation of the evaporation time is based on a QFT on curved background computation, for which back-reactions are neglected.
		It is thus important that the spacetime curvature at the horizon remains sub-Planckian which is not the case when $z_f$ becomes of order one\footnote{A more detailed analysis of the relevance of back-reaction effects is discussed in \cite{NicoliniHolographicScreensin}. There, the ratio $M_{ext}^2 \,T(M_{max})/M_{max}$ of the maximal temperature $T(M_{max})$ and the mass value $M_{max}$ for which this maximum is reached is used to quantify the amount of back-reactions. Comparing to Fig.~\ref{Temperaturegraph}, we find $M_{ext}^2 \,T(M_{max})/M_{max} \simeq 0.04$, which is small and back-reactions should be negligible. Nevertheless, a detailed analysis taking quantum fluctuations into account is still missing for the present model.}.
		The second approximation done here is that it is assumed that the quantum states of spacetime characterize an effective spacetime, where quantum fluctuations can be neglected.
		Whether this is true for small masses, i.e. $z_f \sim 1$, is far from being clear remaining at a pure effective level and the analysis of the proper quantum model, its Hilbert space, and semi-classical sector would be required.
		Quantum effects could change the picture dramatically as they might allow for black-to-white hole tunneling as discussed e.g. in \cite{BianchiWhiteHolesas,HaggardBlackHoleFireworks,Martin-DussaudEvaporationgblackto,DAmbrosioTheEndof,ChristodoulouCharacteristicTimeScales}.
		
		\section{Discussion and Outlook}\label{Sec:conclusion}
		In this paper, we studied the relevant thermodynamic quantities, i.e. temperature $T$, specific heat $C$, and entropy $S$ of the quantum corrected polymer black hole model \cite{BodendorferMassandHorizon,Bodendorferbvtypevariables}.
		It is important to note that here the semi-classical thermodynamic relations for black holes were used.
		At this stage, it is assumed that these remain correct even if quantum gravity corrections are taken into account.
		It would be necessary to analyze the stability of the quantum corrected spacetime against perturbations and check that a generalized version of the no-hair theorem holds, i.e. a quantum black hole can still be described by only a finite set of macroscopic parameters, say the black- and white hole mass for the case of a non-charged and non-rotating black hole considered here.
		A detailed analysis, on the one hand, would require a full (effective) theory of generic charged and rotating quantum black holes, which is currently out of reach (see however \cite{CaravelliSpinningLBH,LiuShadowAndQuasinormal, BrahmaObservationalConsequences} for some progress using the Newman-Janis algorithm). On the other hand, remaining at the level of effective Schwarzschild geometry, it would also be interesting to systematically study the role of asymptotic symmetry charges in the thermodynamic analysis and the role of the second mass observable (WH resp. BH) from the perspective of the other asymptotic spacetime region (BH resp. WH) where only the first observable can be related to the ADM mass.
		The computation of the standard thermodynamic quantities as determined by the usual laws of black hole mechanics, now specified for the effective metric describing our quantum corrected black hole spacetime,  leads to the quantum corrected black hole thermodynamics presented in this work.
		In Fig.~\ref{Temperaturegraph}, the black hole temperature as a function of the black hole mass for the mass relation Eq.~\eqref{eq:quadraticrelation} ($M_{WH} \propto M_{BH}^2$) is shown.
		The qualitative behavior of the results remains unaltered by the specific choice of the mass relation which we left in fact unspecified as much as possible in our analysis.
		It is clearly visible that the temperature approaches the classical result for large black hole masses ($M_{BH} M_{WH} \gg \lambda_k$).
		For smaller masses, the temperature reaches a maximal value and the black hole starts to cool further down as it losses mass.
		This continues up to the extremal point ($M_{BH} M_{WH} = \lambda_k$), where black and white hole horizons coincide.
		The temperature of this configuration is zero and the radiation and evaporation process stops -- the classical divergence is avoided. 
		The details of the temperature maximum (its value and the mass for which it occurs) depend on the mass relation $M_{WH}(M_{BH})$, while the vanishing of the temperature at the extremal configuration and the classical limit being approached for large masses are independent thereof. From the temperature follows directly the specific heat, which is again shown in Fig.~\ref{Ccomparison} for the quadratic mass relation \eqref{eq:quadraticrelation}.
		Again, for large masses, the classical behavior is approached, independently of the mass relation.
		The maximal value of the temperature becomes visible in the specific heat as a divergence which indicates the occurrence of a phase transition.
		As before, the specific quantitative details depend on the concrete mass relation, while its existence is generic. For masses above this critical value, the specific heat is negative, i.e. the black hole is heating up while it loses energy.
		After the phase transition, the specific heat is positive and the black hole cools down as it losses mass.
		This is compatible with the observation from other quantum corrected black hole models (see e.g. \cite{ModestoSelf-dualBlackHoles,NicoliniTdualBH}).
		The model shares with \cite{NicoliniTdualBH} the feature that the specific heat approaches zero when the extremal configuration is reached.
		This is converse to other LQG-inspired models as \cite{ModestoSelf-dualBlackHoles}, where instead the specific heat diverges for small masses.
		As the specific heat reaches zero, the black hole forms (asymptotically in time) a thermodynamically stable remnant with the vanishing temperature at the extremal configuration, where both horizons coincide. Finally, according to the first law of black hole thermodynamics, integrating the inverse temperature w.r.t. the black hole mass gives the entropy. The result for the mass relation \eqref{eq:quadraticrelation} is shown in Fig.~\ref{BHentropyplot}.
		Independently of the specific mass relation, the classical area law $S = A/4$ is recovered for large masses. For the extremal configuration, instead, the entropy vanishes by definition.
		
		Although this gives a regular picture of the semi-classical black hole evaporation process it has to be emphasized that one should not take these results too seriously as the computation is done within the effective approximation.
		Besides the limitations of symmetry reduced quantization, here it is assumed that there exists always proper quantum states which remain sharply peaked on the effective spacetime.
		Therefore, quantum properties such as fluctuations and superpositions are systematically neglected from the beginning.
		While it is reasonable to assume that this effective approximation is good for large masses, the question of whether the same holds true for small masses where the curvature of the black hole horizon lies already in the purely quantum regime remains open.
		In order to make a more precise statement about the quality of this approximation and to analyze the consequences of quantum fluctuations of the geometry, the study of the proper quantum theory of the symmetry-reduced model would be needed. The first steps of its construction were already discussed in \cite{BodendorferEffectiveQuantumExtended}.
		
		More certain is then the behavior of the leading order corrections to the thermodynamic quantities whose generic large mass expansions are given in Eqs. \eqref{Tlargemassexpansion}, \eqref{eq:Cexp}, \eqref{entropymassexp} with no mass relation being chosen. The specific form of the corrections depends on the particular relation between the black and white hole masses and one can then ask what kind of phenomenology would correspond to a given mass relation.
		Of special interest in this sense is the correction to the entropy, whose functional dependence on the black hole mass depends directly on the mass relation (cfr. Eq.~\eqref{entropymassexp}).
		It is then obvious that it is possible to obtain any functional form of the first-order entropy correction by choosing the mass relation accordingly.
		Nevertheless, higher-order corrections disagree as these are uniquely fixed by the mass relation. 
		Of particular interest -- as they were determined by numerous different approaches -- are logarithmic entropy corrections.
		For the model studied here, a (negative) logarithmic correction to the entropy can be achieved for the case of the quadratic mass relation Eq.~\eqref{eq:quadraticrelation} which in this sense seems to be thermodynamically preferred and special compared to other mass relations.
		
		Remarkably, as argued in Sec.~\ref{sec:effectiveQT}, this quadratic mass relation is directly related to proper finite-length effects. In this case, indeed, the canonical momentum $j$ is directly related to the extrinsic curvature or an inverse length/areal radius, say $j\sim 1/b$ with truly mass-independent proportionality factors (cfr. Eq.~\eqref{eq:classregimeb+quadratic}).
		This suggests a relation between inverse length quantum corrections and logarithmic entropy corrections which we also argued to be in agreement with the large mass corrections expected from a minimal length generalized uncertainty principle with minimally allowed position uncertainty given by the critical length scale for the onset of quantum effect, the latter being related to the extremal Planckian BH configuration.
		
		Whether something more fundamental may be hidden behind this result is not yet clear to us and further investigations are necessary.
		In this respect, as already mentioned above, a better understanding of the deep quantum regime as well as strengthening the connections with similar results obtained from other approaches may shed some light on their physical interpretation.
		In \cite{NicoliniHolographicScreensin}, for instance, the extremal critical configuration has been interpreted as a Planckian holographic screen which in turn provides the fundamental quantum of area entering the expression of a discretized mass spectrum.
		It would be interesting to check whether such a viewpoint can be implemented in our setting and if the outcome would be compatible with the spectrum of the mass observable resulting from the quantization of the present model.
		To this aim, a full understanding of the quantum theory underlying our effective model would be necessary which is left for future research.
		
		Note that the present analysis based on \cite{Bodendorferbvtypevariables,BodendorferMassandHorizon} together with the methods described in \cite{MuenchEffectiveQuantumDust,BenAchourBouncingcompactobjectsI,BenAchourBouncingcompactobjectsII,BenAchourConsistentblackto} would in principle allow to construct an effective picture of the full collapse and evaporation process of a quantum black hole.
		Again, it would be important to analyze the quantum theory first, to see the range of validity of the effective picture.
		The construction of the symmetry-reduced quantum theory and the merging of all these approaches into a consistent framework is left for future work.
		
		As the last point, let us emphasize that the presented model has also a white hole horizon.
		It would be possible to repeat the thermodynamic analysis also for this second horizon.
		As already remarked in Sec. \ref{Sec:BHtherm}, the model is symmetric under the exchange $M_{BH} \mapsto M_{WH}$ and $x \mapsto -x$.
		Therefore, the thermodynamics of the white hole side would be exactly equivalent, just with the exchange of black and white hole mass in each formula. In particular, for any given mass amplification or de-amplification, we expect the WH thermodynamics to match the results of the BH side for the corresponding inverse mass de-amplification or amplification.

		\appendix
		\section{Near-Horizon Geometry and Temperature}\label{App:nearhorizon&Euclidean}
		
		As a complement to the analysis of the black hole temperature presented in Sec. \ref{Sec:quantcorrectedtemperature}, in this appendix we briefly discuss the near-horizon limit of the effective geometry and the corresponding derivation of the temperature of the Killing horizon via Euclidean methods.
		Starting from the line element \eqref{temp1}, that we recall here for simplicity
		\be\label{App:lineelement}
		\dd s^2=-\left(\frac{8 \lambda_k M_{BH}^2}{M_{WH}}\right)^{\frac{2}{3}}\frac{a(\tilde{r})}{\lambda_j^2}\dd\tau^2+\frac{\lambda_j^2}{a(\tilde{r})} \left(\frac{8 \lambda_k M_{BH}^2}{M_{WH}}\right)^{-\frac{2}{3}}\dd\tilde{r}^2+b^2(\tilde{r})\dd\Omega_2^2\;,
		\ee
		\noindent
		the near-horizon limit can be studied by considering an expansion around $\tilde{r}=\tilde{r}_++\eta$ with $0<\eta\ll\tilde{r}_+$ (just outside the horizon) so that, at first order in $\eta$, we have
		\begin{align}
		a(\tilde{r}_++\eta)&\simeq a(\tilde{r}_+)+\eta\frac{\dd a(\tilde r)}{\dd\tilde r}\biggl|_{\tilde r=\tilde{r}_+}=\eta\frac{\dd a(\tilde r)}{\dd\tilde r}\biggl|_{\tilde r=\tilde{r}_+}\;,\\
		b^2(\tilde{r}_++\eta)&\simeq b^2(\tilde{r}_+)+2\eta b(\tilde{r}_+)\frac{\dd b(\tilde r)}{\dd\tilde r}\biggl|_{\tilde r=\tilde{r}_+}=b^2(\tilde{r}_+)+\mathcal O(\eta)\;,
		\end{align}
		\noindent
		and the line element \eqref{App:lineelement} decomposes into the sum of a 2-sphere and a 1+1 Lorentzian geometry. The latter reads as
		\be\label{App:2dmetric}
		\dd s_2^2=-\left(\frac{8 \lambda_k M_{BH}^2}{M_{WH}}\right)^{\frac{2}{3}}\frac{\eta}{\lambda_j^2}\frac{\dd a(\tilde r)}{\dd\tilde r}\Bigl|_{\tilde r=\tilde{r}_+}\dd\tau^2+\frac{\lambda_j^2}{\eta\frac{\dd a(\tilde r)}{\dd\tilde r}\bigl|_{\tilde r=\tilde{r}_+}} \left(\frac{8 \lambda_k M_{BH}^2}{M_{WH}}\right)^{-\frac{2}{3}}\dd\eta^2\;.
		\ee
		Performing now the change of variables
		\be\label{App:change2Rindler}
		\eta\longmapsto\rho^2=\frac{4\lambda_j^2}{\frac{\dd a(\tilde r)}{\dd\tilde r}\bigl|_{\tilde r=\tilde{r}_+}} \left(\frac{8 \lambda_k M_{BH}^2}{M_{WH}}\right)^{-\frac{2}{3}}\eta\;,
		\ee
		\noindent
		the line element \eqref{App:2dmetric} can be brought into the Rindler form
		\be\label{App:Rindler}
		\dd s_2^2=-\left(\frac{\rho}{2\lambda_j^2}\left(\frac{8 \lambda_k M_{BH}^2}{M_{WH}}\right)^{\frac{2}{3}}\frac{\dd a(\tilde r)}{\dd\tilde r}\biggl|_{\tilde r=\tilde{r}_+}\right)^2\dd\tau^2+\dd\rho^2=-\rho^2\dd\bar\tau^2+\dd\rho^2\;,
		\ee
		\noindent
		where in the second equality we defined $\bar\tau=\frac{\tau}{2\lambda_j^2}\left(\frac{8 \lambda_k M_{BH}^2}{M_{WH}}\right)^{\frac{2}{3}}\left|\frac{\dd a(\tilde r)}{\dd\tilde r}\right|\Bigl|_{\tilde r=\tilde{r}_+}$. The metric \eqref{App:Rindler} describes a near-horizon frame which is accelerating to keep from falling into the black hole horizon. The local acceleration is given by $\alpha=1/\rho$ and diverges as $\rho\to0$ (i.e. $\eta\to0$ being $\frac{\dd a(\tilde r)}{\dd\tilde r}\bigl|_{\tilde r=\tilde{r}_+}\neq0$) as expected. The expressions for the surface gravity and temperature given in Eqs. \eqref{kappabardeenourbh}-\eqref{TBHfinal} can then be obtained from the local Unruh temperature experienced by such a near-horizon observer, $T(\tilde r)=\frac{1}{2\pi\rho(\tilde{r})}$ with $\rho(\tilde{r})$ given by the inversion of \eqref{App:change2Rindler} together with the definition $\eta=\tilde{r}-\tilde{r}_+$, and taking into account the gravitational redshift at infinity, namely $T=T(+\infty)=\lim_{\tilde{r}\to\tilde{r}_+}T(\tilde{r})\sqrt{\frac{g_{\tau\tau}(\tilde{r})}{g_{\tau\tau}(+\infty)}}$ with $g_{\tau\tau}$ given in \eqref{App:lineelement}.
		Equivalently, going back to the near-horizon Rindler metric \eqref{App:Rindler}, we can put it into the Minkowski form $\dd s_2^2=-\dd\mathfrak{T}^2+\dd\mathfrak{X}^2$ by means of the standard coordinate redefinition $\mathfrak{T}=\rho\sinh(\bar\tau)$, $\mathfrak{X}=\rho\cosh(\bar\tau)$. Therefore, after a Wick rotation to the Euclidean time $\bar{\tau}_E=i\bar{\tau}$ (i.e. $\tau_E=i\tau$), the $2\pi$-periodicity of $\mathfrak{T}_E=\mathfrak{T}(\bar{\tau}_E)$ implies the periodicity of $\tau_E$ to be (cfr. below Eq. \eqref{App:Rindler})
		\be\label{Euclperiodicity}
		\beta=\frac{2\pi}{\frac{1}{2\lambda_j^2}\left(\frac{8 \lambda_k M_{BH}^2}{M_{WH}}\right)^{\frac{2}{3}}\left|\frac{\dd a(\tilde r)}{\dd\tilde r}\right|\Bigl|_{\tilde r=\tilde{r}_+}}\;,
		\ee
		\noindent
		which is compatible with the requirement of absence of conical singularities at $\tilde{r}=\tilde{r}_+$ for the Euclidean metric obtained from \eqref{App:lineelement} after Wick rotation as can be seen by focusing on the metric $\dd s_E^2$ in the $(\tau_E,\tilde{r})$-plane, defining $R=\left(\frac{8 \lambda_k M_{BH}^2}{M_{WH}}\right)^{\frac{1}{3}}\sqrt{\frac{a(\tilde{r})}{\lambda_j^2}}$ so that
		\be
		\dd s_E^2=R^2\dd\tau_E^2+4\lambda_j^4\left(\frac{8 \lambda_k M_{BH}^2}{M_{WH}}\right)^{-\frac{4}{3}}\left(\frac{\dd a(\tilde r)}{\dd\tilde r}\right)^{-2}\dd R^2\;,
		\ee
		\noindent
		and demanding the ratio of the circumference of the circle of radius $\delta R$ and its radius to be $2\pi$ in the $\tilde r\to\tilde{r}_+$ ($R\to0$) limit \cite{FullingTemperaturePeriodicityHorizons,AshtekarPropertiesofarecent,FaraoniUnsettlingphysicsin,Bouhmadi-LopezAsymptoticnon-flatness}. Finally, the expression of the temperature $T=1/\beta$ with $\beta$ given in Eq. \eqref{Euclperiodicity} matches with the results \eqref{kappabardeenourbh}-\eqref{TBHfinal} obtained in Sec. \ref{Sec:quantcorrectedtemperature}.

	\section*{Acknowledgements}
		
		The authors would like to thank Norbert Bodendorfer for valuable discussions and feedback.  The work of SP is supported by an International Junior Research Group grant of the Elite Network of Bavaria. FMM and JM acknowledge hospitality at the University of Regensburg and the support of the Elite Network of Bavaria in the early stages of the project. The work of FMM was supported in part by funding from Okinawa Institute of Science and Technology Graduate University.
		The work of JM at the University of Marseille was made possible through the support of the ID\# 61466 grant from the John Templeton Foundation, as part of the \textit{The Quantum Information Structure of Spacetime (QISS)} Project (qiss.fr). The opinions expressed in this publication are those of the author(s) and do not necessarily reflect the views of the John Templeton Foundation.

	\end{document}